\documentclass{ieeeaccess}

\usepackage{amsmath}
\usepackage{color}
\usepackage{array}
\usepackage{stfloats}
\usepackage{amsthm} 
\usepackage{amssymb}
\usepackage{float}
\usepackage[normalem]{ulem}
\usepackage{soul}
\usepackage{nccmath}
\usepackage{graphicx} 
\usepackage{setspace}
\usepackage{booktabs}
\usepackage{subcaption}
\usepackage{mwe}
\usepackage{cite}
\usepackage{amsmath,amssymb,amsfonts}
\usepackage{textcomp}
\usepackage{color}
\usepackage{colortbl}
\usepackage[utf8]{inputenc} % use UTF8 encoding
\usepackage{kotex} % use KoTeX package for Korean 

\usepackage{algorithm}
\usepackage[noend]{algpseudocode}
\usepackage{algorithmicx}
\usepackage{grffile}
\usepackage{multirow}
\usepackage{multicol}
\usepackage{mathtools}

\usepackage{xspace}
\def\BibTeX{{\rm B\kern-.05em{\sc i\kern-.025em b}\kern-.08em
    T\kern-.1667em\lower.7ex\hbox{E}\kern-.125emX}}

\begin{document}

\history{Date of publication xxxx 00, 0000, date of current version xxxx 00, 0000.}
\doi{10.1109/ACCESS.2017.DOI}
\title{Workload-Aware Scheduling using Markov Decision Process for Infrastructure-Assisted Learning-Based Multi-UAV Surveillance Networks}

\author{\uppercase{Soohyun Park}\authorrefmark{1}, 
\uppercase{Chanyoung Park}\authorrefmark{1},
\uppercase{Soyi Jung}\authorrefmark{2} \IEEEmembership{Member, IEEE},
\uppercase{Jae-Hyun Kim}\authorrefmark{2} \IEEEmembership{Member, IEEE}, and
\uppercase{Joongheon Kim}\authorrefmark{1} \IEEEmembership{Senior Member, IEEE}}
\address[1]{Department of Electrical and Computer Engineering, Korea University, Seoul, Korea (e-mails: soohyun828@korea.ac.kr, cosdeneb@korea.ac.kr)}
\address[2]{Department of Electrical and Computer Engineering, Ajou University, Suwon, Korea}
\tfootnote{The authors acknowledge the support from Nano UAV Intelligence Systems Research Laboratory at Kwangwoon University, funded by Defense Acquisition Program Administration (DAPA) and Agency for Defense Development (ADD) (UD200027ED).}

\markboth
{S. Park \headeretal: Workload-Aware Scheduling using MDP for Infrastructure-Assisted Learning-Based Multi-UAV Surveillance Networks}
{S. Park \headeretal: Workload-Aware Scheduling using MDP for Infrastructure-Assisted Learning-Based Multi-UAV Surveillance Networks}

\corresp{Corresponding authors: Soyi Jung, Jae-Hyun Kim, and Joongheon Kim (e-mails: sjung@ajou.ac.kr, jkim@ajou.ac.kr, joongheon@korea.ac.kr).}

\begin{abstract}
In modern networking research, infrastructure-assisted unmanned autonomous vehicles (UAVs) are actively considered for real-time learning-based surveillance and aerial data-delivery under unexpected 3D free mobility and coordination. 
In this system model, it is essential to consider the power limitation in UAVs and autonomous object recognition (for abnormal behavior detection) deep learning performance in infrastructure/towers. 
To overcome the power limitation of UAVs, this paper proposes a novel aerial scheduling algorithm between multi-UAVs and multi-towers where the towers conduct wireless power transfer toward UAVs.
In addition, to take care of the high-performance learning model training in towers, we also propose a data delivery scheme which makes UAVs deliver the training data to the towers fairly to prevent problems due to data imbalance (e.g., huge computation overhead caused by larger data delivery or overfitting from less data delivery). Therefore, this paper proposes a novel workload-aware scheduling algorithm between multi-towers and multi-UAVs for joint power-charging from towers to their associated UAVs and training data delivery from UAVs to their associated towers. To compute the workload-aware optimal scheduling decisions in each unit time, our solution approach for the given scheduling problem is designed based on Markov decision process (MDP) to deal with (i) time-varying low-complexity computation and (ii) pseudo-polynomial optimality. As shown in performance evaluation results, our proposed algorithm ensures (i) sufficient times for resource exchanges between towers and UAVs, (ii) the most even and uniform data collection during the processes compared to the other algorithms, and (iii) the performance of all towers convergence to optimal levels.
\end{abstract}


\titlepgskip=-15pt

\maketitle
\section{Introduction}\label{sec:intro} 
\subsection{Background and Motivation}
In various surveillance situations where it is necessary to monitor, manage, and detect damages and abnormal behaviors in large areas, unmanned aerial vehicle (UAV) systems have been widely and actively used as one of emerging and representative solutions~\cite{elec[55], elec[56], elec[57], tiiyun}.
The UAV surveillance system produces and transmits environment observation data to system infrastructure (e.g., towers and base stations) in an ad-hoc and flexible manner~\cite{access[24],access[25], access[27]}. 
According to the properties of high mobility and free arrangement in UAV networks, the UAV devices can observe and collect information in the environment where it is burdensome to design and plan new ground infrastructure networks~\cite{access, electronics}.
For this reason, in particular, the UAV surveillance system is suitable for the real-time system safety and utility maintenance of areas which are sensitive to external attacks and damages.
The observed environment data obtained by the multiple UAVs may be transferred to the infrastructures which are around the UAVs to be used as training data in order to build autonomous computer vision based surveillance deep learning models (especially for object detection models)~\cite{elec[1],elec[2], elec[53], elec[54]}. Therefore, it is obvious that \textit{workload-aware} algorithms are required for UAV-based autonomous surveillance networks where the workload can be defined as the number of training data which are obtained by UAVs and processed at the scheduled/associated infrastructure.
The performances (e.g., accuracy) of the autonomous surveillance deep learning models are affected by the size and characteristics of the obtained data by multiple UAVs~\cite{sensors[12], sensors[13],sensors}, thus, appropriate scheduling/matching algorithms between UAVs and infrastructure stations are essentially required under the consideration of workload~\cite{elec[5], access[3], access[16], access[20]}.
Therefore, from a data transmission perspective, one of the important considerations of the scheduling/matching methods in multi-UAV surveillance networks is the data retention status of infrastructure (i.e., towers which act as base stations, the data collectors from UAVs, and power-charging facilities for wireless power transfer to UAVs)~\cite{access[6], access[33], elec[49]}. 

Moreover, according to the fact that UAV devices are power hungry, it is essential to consider energy-efficiency and power consumption for designing and implementing UAV networks~\cite{access[1], access[24], access[29], access[4]}. Traditionally, in order to address UAV power issues, the scenarios have always been assumed and considered in which UAVs under charging services via energy-rich system infrastructure (acting as a power-charging facilities)~\cite{access, access[1], access[2],access[3], tvtjung}. Recently, there have been active studies that extend the operation time of multi-UAV networks by providing sufficient power-sources to UAVs by utilizing mobile charging stations via vehicles or heterogeneous UAVs~\cite{access[8], access[9], access[10], access[11]}.
As many studies have already shown and discussed, scheduling algorithms for power transmission in multi-UAV networks ensure the stability of the multi-UAV networks as well as extend the valid system operating times in order to achieve system objectives using multiple UAVs~\cite{elec[57], access[29], elec[45], access}. 

\begin{figure*}
    \centering
    \includegraphics[width=0.911\linewidth]{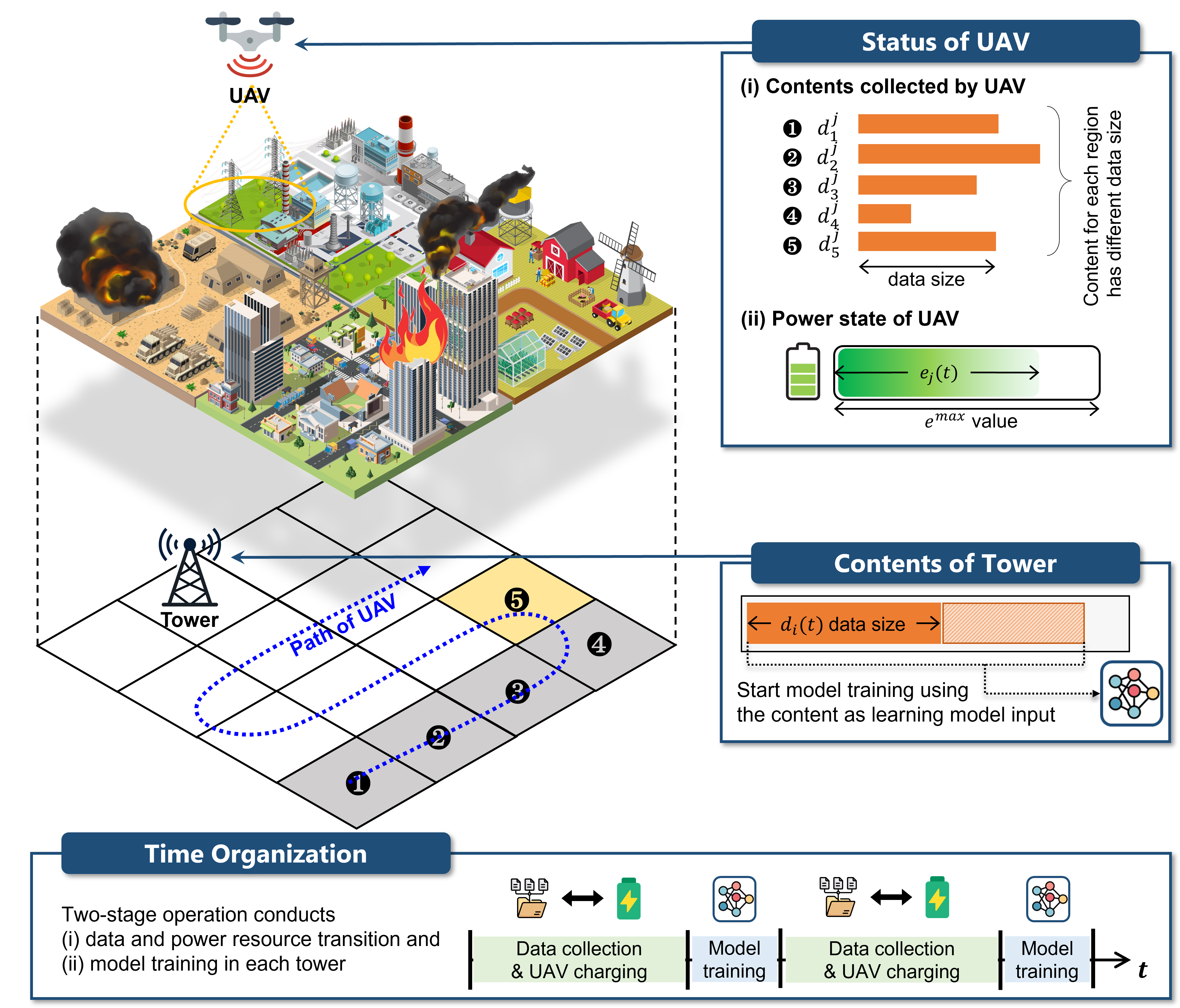}
    \caption{Overall system architecture for infrastructure-assisted multi-UAV surveillance networks.}
    \label{fig:overall_architecture}
\end{figure*}

This paper considers the scenario for real-time autonomous surveillance in extreme large-scale areas using multiple UAVs and their supporting infrastructure (i.e., multiple towers) under the consideration of UAV characteristics.
Here, the towers train a well-trained object (damages and abnormal behaviors) recognition model where the damages and abnormal behaviors are caused by unexpected sudden external attacks (e.g., fires or smokes).
In addition, in the system, UAVs provide the training data which are used for autonomous object recognition deep learning model training in each tower and the tower acts as a charging station because the tower always has sufficient power resources in order to charge scheduled/associated UAV devices that come to transfer collected data. 
In other words, this paper considers a system model in which the UAVs and their associated towers exchange necessary resources~\cite{access[16], elec[5]} by receiving power-charging from the associated tower while delivering data held by the associated UAV, as shown in Fig.~\ref{fig:overall_architecture}.
Moreover, in the environment where the area is sensitive to external attacks or damages, consideration of the time limit is also important.
In this research, the time limit situations stand for mission-critical situations in which the time allowed to collect data using the multi-UAV devices are limited due to any damages, or when all towers train the models, the time for completing the learning may not be guaranteed.
Furthermore, the environment area is divided into multiple regions in the form of a grid map shown in Fig.~\ref{fig:overall_architecture}. It is assumed that the damages such as fire alarms or smokes occur randomly in each region, for certain regions, a larger amount of observation data can be created than the other regions.

When the surveillance UAV devices are flying over the region, each UAV device converts the observed data into one content. The content generated for each region is stored in its own storage as many as the number of regions passed by each UAV along the predefined trajectory~\cite{electronics}.
When the towers perform surveillance object recognition deep learning model training, the data size held in each tower impacts the time required to complete learning~\cite{sensors, sensors[12]}. 
In our considering system, if there is an imbalance in the size of the data collected in towers, the towers that occupy a lot of data spend too much time for model training. In addition, it results in deviations in the performance of the learning model of the towers.
On the other hand, if towers have insufficient size of data for model training, the performance of the model will be degraded due to overfitting.
Therefore, it is difficult to say that stable and reliable model training is progressed in the system. Therefore, workload-aware scheduling between multi-UAVs and towers should be designed and implemented.

For the design and implementation of the workload-aware scheduling algorithm, the proposed algorithm in this paper divides the system operation time into two slots/periods, i.e., (i) the period of data collection \& UAV charging and (ii) the period of performing learning with the collected data. The first period is for the workload-aware scheduling to ensure that all towers existing in the system are guaranteed to have high-accurate autonomous object detection deep learning models without significant deviation.
The purpose of our research is to design and implement a novel workload-aware scheduling algorithm which ensures that (i) all towers in the system are provided with fair data, guarantees uniform and non-biased performance after collecting a certain amount of data; and (ii) a multi-UAV based system has a sufficient retention time to reliably and robustly perform the missions by receiving power-charging via wireless power transfer from the associated/scheduled towers.
In order to guarantee the robust and reliable operations of infrastructure/tower-assisted multi-UAV surveillance networks, our proposed scheduling algorithm design and implementation should be based on optimization framework for maximizing following two objectives, i.e., \textit{(Objective 1)} the accumulated learning/training data of each tower under the consideration of data size fairness and \textit{(Objective 2)} the average UAVs power amounts which can guarantee system retention (based on energy-efficiency) during the period for data and power resources exchanges. 
In order to achieve our desired goals (the consideration of \textit{(Objective 1)} and \textit{(Objective 2)}), a novel workload-aware scheduling algorithm is designed under the consideration of tower and UAV conditions. The proposed solution approach for the proposed optimization formation is built based on Markov decision process (MDP) which is one of major reinforcement learning algorithms that is widely used because it is mathematically trackable and analyzable. Based on the MDP-based workload-aware scheduling algorithm design, discrete-time stochastic control via sequential decision making can be available and it is beneficial as follows.
\begin{itemize}
    \item As time changes in our time-dependent optimization formulation, calculating the optimal solution continuously and iteratively is a huge burden in a situation where the number of towers and UAVs is larger. Furthermore, the number of UAVs can be varied due to the power status of UAVs. Therefore, MDP-based discrete-time sequential decision making for scheduling action decision is practical and computationally beneficial.
    \item If the scheduling decisions are made in each unit time using MDP-based approach, optimal solutions can be obtained with pseudo polynomial time~\cite{twc201912choi}. This is a big deal because conventional scheduling problems are generally integer problems which is one of well-known combinatorics problems (NP-Hard). Therefore, it is obvious that our proposed MDP-based workload-aware scheduling algorithm which works as discrete-time sequential decision making is the best approach. 
\end{itemize}

\subsection{Contributions}
The main contributions of this research are as follows.
\begin{itemize}
    \item To establish a safety management system that can detect random accidents or damages in real-time, we propose a system that ensures an even level of all autonomous high-performance object detection deep learning models at infrastructure/towers in multi-UAV surveillance networks. For achieving the even level of performance, the proposed scheduling between towers and UAVs is for even/fair training data distribution from UAVs to their associated towers. Therefore, the proposed scheduling algorithm is for workload-balanced (i.e., workload-aware) at towers and also for power-consumption-considered at UAVs. 
    \item In order to solve the workload-ware scheduling optimization, MDP-based approach is designed and implemented for finding optimal solutions in each time step to exchange necessary resources between multi-tower and multi-UAV. The use of MDP can lead to optimal solutions for time-dependent optimization formulation in time-varying dynamic networks. Through this, the sufficient operating time due to low-computation via MDP-based approach and the learning model accuracy of UAV-based systems are guaranteed.
    \item The performance of the proposed algorithm is evaluated and analyzed in various ways by considering the characteristics of the actual UAV movement models and also by conducting experiments using different real-world learning models/datasets that can be learned and utilized in towers.
\end{itemize}

\subsection{Organization}
The rest of this paper is organized as follows.
Sec.~\ref{sec:2} presents the preliminary knowledge, i.e., reference network architecture and UAV model/mobility models. Sec.~\ref{sec:3} describes the details of the proposed MDP-based scheduling algorithm between UAVs and towers for workload-fair contents access. Sec.~\ref{sec:4} evaluates and analyzes the performance of the proposed algorithm, and finally, Sec.~\ref{sec:5} concludes this paper.

\section{Preliminaries}\label{sec:2}
The related work in this research is summarized in Sec.~\ref{sec:related}. In addition, our considering reference network architecture is well-described in Sec.~\ref{sec:2-1}. After that the models of UAV-specific mobility and power/energy are presented in Sec.~\ref{sec:2-2} and Sec.~\ref{sec:2-3}, respectively.

%In this section, we describe the system model where the proposed content access algorithm is applied. The following subsections describe the network model and UAVs' movement and power model. We assume that there are no transmission failure factors such as packet loss and interference when the UAVs transmit several contents to the matched tower.

\subsection{Related Work}\label{sec:related}
The network organization and coordination using multiple UAVs devices (i.e., multi-UAV networks) has various advantages due to the mobility of UAV devices. However, according to the existence of uncertainties in real-world environment, application-specific and optimal control schemes are vital for multi-UAV networks in order to achieve the desired goals. For constructing autonomous multi-UAV network systems, the research result in~\cite{jung2021adaptive} proposes a new scheduling algorithm for surveillance in UAV-assisted smart city platforms~\cite{9548768}. In~\cite{jung2021adaptive}, all UAVs in this system collect image information and transmit it to multi-access edge computing (MEC) systems for stabilized super-resolution application in order to realize robust and reliable autonomous surveillance. However, each UAV's energy state needs to be considered to establish UAV-based platforms in real-world environment.
Moreover, there are various studies~\cite{access[33], elec[49], access[16], access[18], access[28]} that consider the energy issues in UAVs for the scenarios where UAVs transmit data to their associated infrastructure.
In~\cite{access[33]}, a new algorithm is proposed that is for UAV power consumption minimization during data collection. In situations where the UAV's power is limited without additional power-charging considerations, non-convex optimization formulation is established and the formulation is solved via an iteration-based algorithm. In~\cite{access[33]}, the authors insist that the proposed method greatly minimizes the power consumption of UAVs proportional to the transmitted data size than other algorithms. Another research to minimize UAVs' power consumption is for extending the network lifetime with UAVs as mobile data collectors. In~\cite{elec[49]}, the proposed objective function for the energy-efficiency of UAVs is computed using successive convex optimization. The authors in~\cite{elec[49]} guarantee that the required size of data is collected reliably by minimizing the power consumption through the optimized UAV trajectories. 
Moreover, several research results optimize power efficiency and data collection utility in multi-UAV networks. Most researches adaptively control the power consumption of UAVs themselves rather than additional power supply through charging for data transmission. 
The proposed algorithm in \cite{wang2021federated} performs UAV-assisted crowd-sensing to facilitate federated learning (FL) services. It proposes the fair deployment scheme of edge computing devices at multiple small and conventional macro cell base stations for efficient data training and model exchange. By designing an optimal incentive mechanism to encourage UAVs' participation in FL fairly, they improve the user utility as well as communication efficiency compared to existing FL schemes. Therefore, a fair data distribution is essential in training artificial intelligence (AI) models to improve the performance of the overall network systems.

\subsection{Reference Network Architecture}\label{sec:2-1}

\subsubsection{Surveillance Network Segmentation}
As illustrated in Fig.~\ref{fig:overall_architecture}, $N$ towers and $M$ UAVs exist in our considering reference surveillance network architecture, where the two types of network components (i.e., towers and UAVs) participate in the resource exchange scheduling. Note that the entire surveillance network area is converted into segmented grid environment which has identical-size $L$ regions. Here, our considering damage/emergency events (e.g., abnormal detection such as fire alarms) happen in independent and identical distribution (i.i.d.) randomly over each region.

% tower
\subsubsection{Towers}
The towers are placed at regular/equivalent intervals on our considering grid map and receive several contents from their associated UAVs that contain various observed data of regions. The towers perform the model training and the purpose of the model is to achieve good surveillance performance by utilizing sufficient amount of data. 
Because each tower is ground-mounted infrastructure, it is assumed that there is no limit to the power supply of the tower. It provides power to scheduled UAVs while receiving contents from the UAVs.
From a tower perspective, because data size is important for conducting high-performance learning tasks, it is important to receive sufficient data from several content files of scheduled UAVs using a limited number of charging panels.

%UAV
\subsubsection{UAVs}
Each UAV collects the region's observation data through the UAV-mounted camera and generates data as one content for the region while it is hovering. The collected/generated content is stored at the memory space of UAV before the UAV is scheduled to a specific tower for delivering the content. When the UAV moves to its associated/scheduled tower, the UAV transmits its own all stored contents, and after that, the UAV moves back to its own trajectory and gathers the region's observation data for contents generation, again. When the UAV and its associated/scheduled tower are close to each other, the data transmission and power-charging occur simultaneously between them. 
If proper scheduling is not made, UAVs continue to consume its own power and collect information along the path. These results can significantly degrade the overall system performance because the UAVs may shut down quickly, etc.

%
%We assume the multi-UAV-based network where fairly content access is necessary in mission-critical environment. 

\subsection{UAV Mobility Model}\label{sec:2-2}
We assume that UAVs fly and hover at same altitude $h$. In addition, the UAVs have the same radius $r$ of surveillance. Therefore, it is obvious that the surveillance area by the UAV $j$ is $a_j = r^2\pi$. Here, the radius $r$ can be calculated as, 
\begin{equation}
r = h\cdot \tan\left(\frac{FoV}{2}\right)
\end{equation}
where $FoV$ is the field of view of the UAV-mounted surveillance camera~\cite{electronics}. 
We also assume that the movement of each UAV is based on the pre-determined trajectory and the trajectory consists of the sequence of several way-points. Therefore, UAVs (i) fly to the next positions from the current positions through the trajectory path; and then (ii) collect surveillance data from the regions where they are located at. When UAVs on their pre-determined paths are scheduled with specific towers, they move from their current positions to their associated/scheduled towers, and then, the UAVs deliver their entire contents to the associated towers. Finally, the UAVs return to their original positions via their own predefined travel paths within one-time step intervals.

\subsection{UAV Power/Energy Model} \label{sec:2-3}

As well-studied in many research results in UAV networks~\cite{tvt201905shin,tvt202106jung}, the consideration of power/energy component is the most essential part in UAV network design and its related optimization. As clearly explained and discussed in~\cite{elec[65]}, the power model of UAVs is divided into two parts, i.e., power acquisition model (via wireless power transfer from scheduled towers) and power consumption model, respectively.

\subsubsection{Power Acquisition Model} 
If UAVs are scheduled with towers, the UAVs fly to the positions of their scheduled towers, and the UAVs receive wireless power transfer charging services from the scheduled towers through the charging panels attached to the towers. During the wireless charging, electromagnetic losses over wireless channels (i.e., $\eta_{i}^{T}$ at towers and $\eta_{j}^{U}$ at UAVs) may occur for the powers provided by scheduled towers (i.e., $\mathcal{E}^{t}_j$). The total amount of charged power/energy (i.e., $e^c$) when $j$-th UAV actually receives after the wireless power transfer charging service is terminated can be calculated as~\eqref{eq:wireless_charging_amount},
\begin{equation}
e^c = \mathcal{E}^{t}_j\cdot \eta_{i}^{T}\cdot\eta_{j}^{U}\cdot \left(u^{t}-\frac{l_{ij}(t)}{v_{j}}\right)     \label{eq:wireless_charging_amount}
\end{equation}
where $u^{t}$ means the operation time for the scheduling between UAVs and towers. In addition, $\frac{l_{ij}}{v_j}$ is the time for moving from $i$-th tower to $j$-th UAV at speed $v_j$~\cite{access}.

The amount of $j$-th UAV power state after charging by $i$-th tower is as follows,
\begin{equation}
    \min\left\{e_{j}(t) + e^c, E_{j}\right\}
    \label{eq:charging2}
\end{equation}
which means the summation of $e^c$ and the $j$-th UAV power states before power charging $e_j(t)$ cannot be greater than the maximum power capacity $E_j$ that UAVs can have~\cite{access}.

\subsubsection{Power Consumption Model} 
Each UAV observes the region for surveillance and generates contents while moving along pre-determined paths, except in situations where it has to move to a tower based on scheduling results. In general, the power consumed during the operation of UAV is divided into two types~\cite{elec[62], elec[63]}, i.e., \textit{hovering} and \textit{cruising}.

For \textit{hovering}, the power consumption can be formulated as follows~\eqref{eq:UAV_hovering_power}, 
\begin{equation}
    P^j_h \triangleq \underbrace{\frac{\delta}{8}\rho sA\Omega^3R^3}_{P_o} +\underbrace {(1+k)\frac{W^{3/2}}{\sqrt{2\rho A}'}}_{P_i}
    \label{eq:UAV_hovering_power}
\end{equation}
where the parameters and variables in this equation are summarized in Table~\ref{tab:UAV_power_parameter}. 

For \textit{cruising}, following formulation is utilized because UAVs move at the same speed without hovering~\eqref{eq:UAV_cruising_power}, i.e., 

\begin{equation}
\begin{split}
    P^j_p(v(t)) \triangleq & \underbrace{P_o\left(1 + \frac{3v_j(t)^2}{U^2_{tip}}\right)}_{\textrm{blade profile}} + \\ 
    & \underbrace{P_i\left(\sqrt{1+\frac{v_j(t)^4}{4v_o^4}} - \frac{v(t)^2}{2v_0^2}\right)^{1/2}}_{\textrm{induced}} + \\
    & \underbrace{\frac{1}{2}d_0 \rho sAv_j(t)^3}_{\textrm{parasite}}
\end{split}
    \label{eq:UAV_cruising_power}
\end{equation}
where the parameters and variables in this equation are also summarized in Table~\ref{tab:UAV_power_parameter}. The specific values for the given parameters and variables in Table~\ref{tab:UAV_power_parameter} are in Table~\ref{tab:parameters_of_uav}, and they are used in our simulation-based data-intensive performance evaluation studies. 

\begin{table}[]
%\small
    \caption{Notations for UAV Power/Energy Model}
    \label{tab:UAV_power_parameter}
    \centering
    \begin{tabular}{c||l}
        \toprule[1.0pt]
        Notation & Description\\
        \midrule[1.0pt]
        $\delta$ & Profile drag coefficient\\
        $\rho$ & Air density\\
        $s$ & Rotor solidity\\
        $A$ & Rotor disc area\\
        $\Omega$ & Blade angular velocity\\
        $R$ & Rotor radius\\
        $k$ & Incremental correction factor to induced power\\
        $W$ & Aircraft weight including battery and propellers\\
        $U_{tip}$ & Tip speed of the rotor blade\\
        $v_o$ & Mean rotor-induced velocity in hovering\\
        \bottomrule[1.0pt] 
    \end{tabular}
\end{table}

\begin{figure*}
    \centering
    \includegraphics[width=0.99\linewidth]{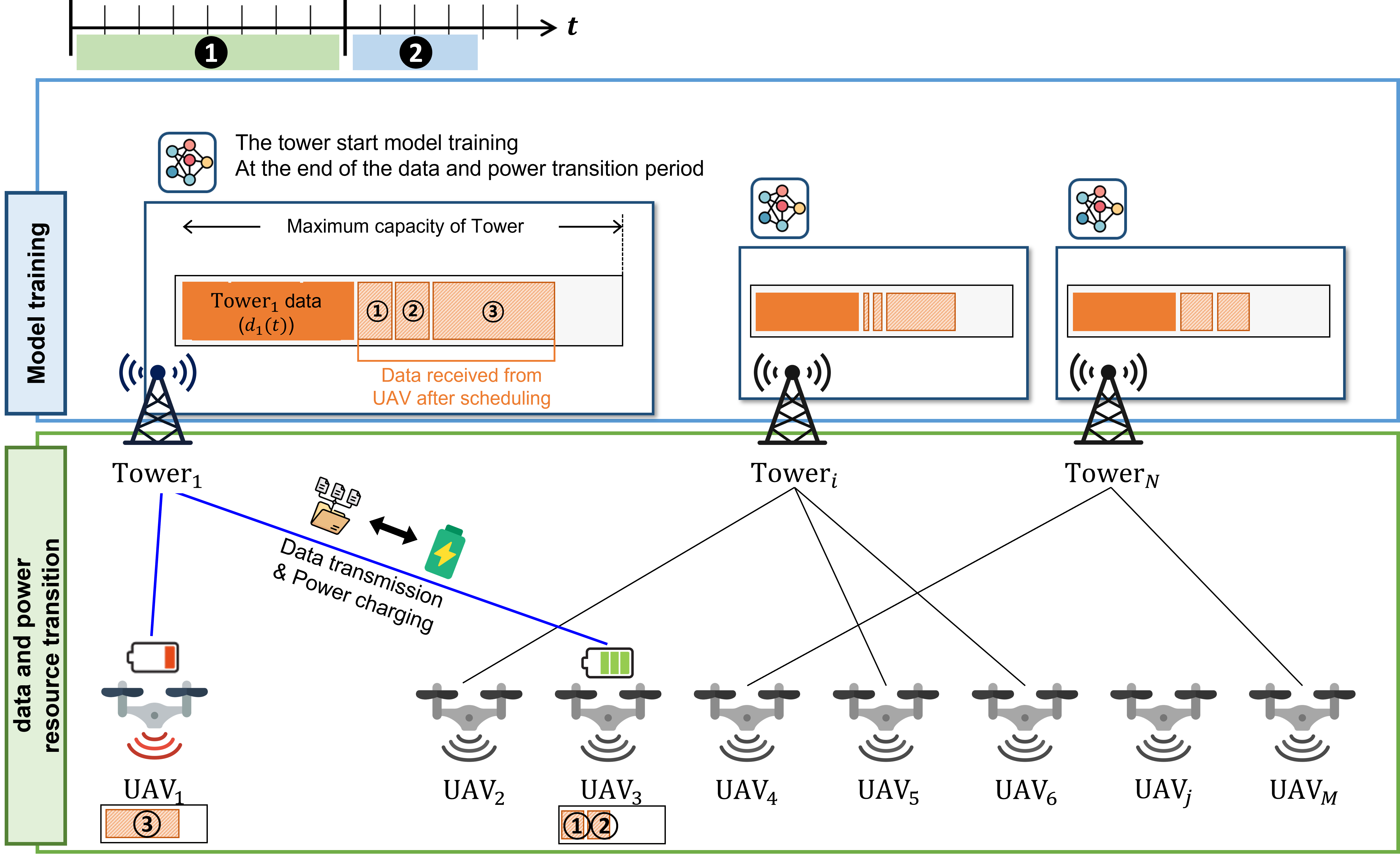}
    \caption{Detailed operations of the proposed scheduling algorithm.}
    \label{fig:period (i)}
\end{figure*}

\section{Workload-Aware Markov Decision Scheduling for Multi-UAV Surveillance Networks}\label{sec:3}

\subsection{Algorithm Design Concept}\label{sec:3-0}
As shown in Fig.~\ref{fig:period (i)}, the towers' data collection is possible only during the data and power resource transition period; and the data only comes from UAVs. Therefore, the proposed workload-aware scheduling algorithm for matching between UAVs and towers is formulated which can ensure that the towers get a lot of data for fair collected data distribution among towers. Note that the UAVs continue the surveillance data collection operation until the end of the period. 

First of all,

\begin{equation}
    \mathcal{V}(t)= \epsilon\cdot \underbrace{{V}^{T}_{data}(t)}_{\text{tower side value}} + (1-\epsilon)\cdot\underbrace{{V}^{U}_{power}(t)}_{\text{UAV side value}}
    \label{eq: utility}
\end{equation}
where ${V}^{T}_{data}(t)$, ${V}^{U}_{power}(t)$, and $\mathcal{V}(t)$ stand for the values which represent the amount of data which the towers fairly have (i.e., tower side value), the average power state of UAVs (i.e., UAV side value), and the degree of optimal safety system management through the two values, respectively. Here, $\epsilon$ means the weight factor between the two values.

Based on the definition of our main objective variable $\mathcal{V}(t)$ in \eqref{eq: utility}, our main objective function can be as follows,
\begin{eqnarray}
    \max: & & \lim_{\mathcal{T}\rightarrow\infty}
    \frac{1}{\mathcal{T}}\sum_{t=1}^{\mathcal{T}}\mathcal{V}(t)
    \label{eq:overall object}
\end{eqnarray}
which means the maximization of time-average $\mathcal{V}(t)$ over $\mathcal{T}$ time steps. Notice that the overall objective which has to be achieved to ensure the system reliability considering the two values (i.e., tower side value and UAV side value) jointly in the proposed network is explained in~\eqref{eq:overall object}.

% 최적화 하는 방법은 왜 MDP?
% 위에서 설계한 목적식 (최대화 문제)를 풀기 위해서 time step마다 optimal solution을 보장하는 MDP를 사용.
% MDP 에 대한 개념 / env <-> agent 교환 (image추가)

% UAV의 이동성 (위치 변화), tower의 상태, UAV가 가지고 있는 content의 가치 (AoI) 변화를 system manager가 관찰하고 목적식을 reward로 하여 MDP 적용 가능.
% agent-> virtual system manager
% action -> # of UAV * # of tower \in 0 or 1 (agent (system manager)에서 결정해야 하는 것 ( == 제안하는 알고리즘의 solution))
% reward -> 위에 설계한 목적식
% 위에 reward maximize를 만족하기 위해 고려되어야 하는 조건들

\subsection{Problem Formulation}\label{sec:3-1}
For robust, reliable, and our application-specific scheduling algorithm design and implementation, (i) the properties of UAVs and (ii) the data size held in towers and UAVs should be considered. The basic optimization formulation for the proposed workload-aware scheduling between UAVs and towers are as follows,

% (i) AoI of tower which is transmitted from UAVs, (ii) data size which is efficient for processing the task, (iii) the amount of energy remaining in each UAV after charging (data transmission). %and (iv) whether to hold content that has exceeded the maximum valid time. 
\begin{eqnarray}
    \max_{x}: & &  \lim_{\mathcal{T}\rightarrow\infty}
    \frac{1}{\mathcal{T}}\sum_{t=1}^{\mathcal{T}}\mathcal{R}(t) \label{eq:objR}\\
    \text{s.t.} & &    
    \mathcal{R}(t) = \epsilon\cdot R^{T}_{data}(t) + (1-\epsilon)\cdot R^{U}_{power}(t) \label{eq: reward}\\
    & & 
    R^{T}_{data}(t) = \frac{\sum^{N}_{i=1}d_i(t)}{\sqrt{\frac{\sum^N_{i=1}(d_i(t) - \Bar{d(t)})^2}{N}}} \label{eq: reward of tower data}\\
    & &
    R^{U}_{power}(t) = \sum^{M}_{j=1}\frac{E_j(t)}{E_j} \label{eq: reward of UAV power}\\
    & &
    d_i(t) = d_i(t-1) + \sum^{M}_{j=1}(\sum^{L^j}_{h=1}d^h_j(t))\cdot x_{ij}(t)\label{eq: tower data calculation}\\
    & &
    E_j(t) = \min\left\{e_{j}(t) + e^c, E_{j}\right\}\cdot x_{ij}(t) + \nonumber\\
    & &\quad\quad\quad\quad e_j(t)\cdot (1-x_{ij}(t)) - P_p(v(t))\label{eq: power calculation}\\
    & &
    \sum^{M}_{j=1}x_{ij}(t) \leq H_i, \forall i \in N \label{eq: matching limit}
\end{eqnarray}
%제안하는 UAV 기반 네트워크 환경에서의 성공적인 content access를 위해 우리는 UAV의 특성과 위에서 정의한 content AoI를 고려해야 한다. tower에서 진행하는 data 기반 learning task의 성능을 보장하고 UAV 기반 네트워크에서의 시스템 유지가 매우 중요. 이러한 이유로 본 section에서는 시스템에서 고려해야하는 4가지 요소를 정의하고 이것들이 혼합된 시스템 utiltiy를 설계하고 이를 최대화하기 위한 optimal scheduling 알고리즘을 제안한다. 
and here, notice that 
    $\mathcal{R}(t)=\mathcal{V}(t)$, 
    $V^{T}_{data}(t)=R^{T}_{data}(t)$, and 
    $V^{U}_{power}(t)=R^{U}_{power}(t)$\footnote{The reason for changing variable names is that the proposed optimization formulation will be solved by MDP-based reinforcement learning which maximizes expected returns. Thus, the three variable notations are updated for better readability.}.
In \eqref{eq:objR}, as explained in~\eqref{eq:overall object}, our main objective is for the maximization of system reliability under the consideration of tower side value $V^{T}_{data}(t)=R^{T}_{data}(t)$ (i.e., fair data distribution for high-performance model training at towers) and UAV side value $V^{U}_{power}(t)=R^{U}_{power}(t)$ (i.e., power consumption minimization at UAVs); and obviously, \eqref{eq: reward} is formulated based on~\eqref{eq:overall object}.
Note that the proposed scheduling algorithm considers the two values equally when $\epsilon=0.5$ in \eqref{eq: reward}.
For more details, $R^{T}_{data}(t)$ is a tower side system value and each tower $i$ has different number of contents (i.e., $L^i$). In \eqref{eq: reward of tower data}, we check the difference between the average data held by towers in the system and the data size held by individual towers. Thus, $R^{T}_{data}$ is the cumulative value for all towers and the larger value means that the towers are with more evenly distributed data profiles.
Moreover, the reward factor for the UAV operation efficiency in the system is formulated as~\eqref{eq: reward of UAV power}. In \eqref{eq: reward of UAV power}, $E_j(t)$ is the energy state of $j$-th UAV after completing the power-charging via its associated/scheduled tower, where the value can be calculated via \eqref{eq: tower data calculation}. 
In \eqref{eq: power calculation}, $e_j(t)$ is the energy state of $j$-th UAV before charging or moving at time step $t$. If $x_{ij}(t)=1$ (i.e., $x_{ij}$ is a scheduling vector between $i$-th tower and $j$-th UAV), the UAV's energy state is changed by~\eqref{eq:charging2} (and there is no additional poser charging when $x_{ij}(t)=0$). $P^j_p(v(t))$ is the power consumed by moving to the next position of the $j$-th UAV. Lastly, \eqref{eq: matching limit} means that there is a scheduling constraint by the number of power-charging panels the tower has (i.e., $H_i$).

% UAV energy 관련 세부 수식은 UAV section 3 reward 부분에서 상세 서술
% Negative reward $R^U_{AoI}(t)$ occurs when UAVs send content that exceeds the maximum allowable time ($A^{max}_h$) to the tower by \eqref{eq: reward of UAV AoI} and \eqref{eq: AoI constraint}. \eqref{eq: data constraint} and \eqref{eq: matching limit} means the constraints of the algorithm. Each represents that all of the towers (i) can process task only when the size of accumulated content received from UAVs is larger than a specific threshold and (ii) can be scheduled at the same time only by the number of panels the tower has ($H_i$).

\subsection{Algorithm Details}\label{sec:3-2}
To solve our proposed optimization formulation for workload-aware scheduling (in Sec.~\ref{sec:3-1}), an MDP-based reinforcement learning model is proposed for the solution approach, i.e., $<\mathcal{S}, \mathcal{A}, \mathcal{P}, \mathcal{R}, \gamma>$ because it can make sequential optimal decisions for time-varying optimization formulation (which can be burden in iterative convex optimization computations in each time step, with large numbers of towers and UAVs) and it can guarantee optimal solutions in pseudo polynomial-time computational complexity even though our considering problem is integer programming~\cite{twc201912choi,6571226,access[23], access[28], access[29], access[31], tvtjung, tiiyun}.
Therefore, in this part, the proposed optimization formulation (i.e., \eqref{eq:objR}--\eqref{eq: matching limit}) designed in~\ref{sec:3-1} is converted into the MDP form as the following subsections.

\subsubsection{State Space}
In our proposed MDP-based scheduling algorithm an agent (i.e., a system manager) schedules multi-UAVs and towers based on the information from the observation of the time-varying network dynamics and the information in the network environment. The information used by the agent represents a state space which contains the status of contents held by towers and UAVs, the distances between of all system components, the energy status values of UAVs, and the channel states of towers, as follows,
\begin{equation}
    \mathcal{S}(t)=\{[d(t)], [l(t)], [h(t)], [e(t)]\}
\end{equation} 
where $[d(t)]$ consists of $d_i(t)$ and $d^j_h(t)$ which are the data sizes of the contents held in $i$-th tower and $j$-th UAV, $[l(t)]$ consists of $l_{ij}(t)$ which is the distance between $i$-th tower and $j$-th UAV, $[h(t)]$ stands for the channel state of towers which is determined by scheduling decision action $x_{ij}(t)$ between $i$-th tower and $j$-th UAV at time $t$, and $[e(t)]$ consists of $e_j(t)$ and $e^f_j(t)$ where the two values mean the UAV's energy retention amount before charging at time $t$ and the amount of energy consumed when $j$-th UAV moves to it associated/scheduled $i$-th tower.

\subsubsection{Action Space} 
In this MDP-based scheduling formulation, the agent's action is replaced as a variable whether the $i$-th tower and $j$-th UAV are scheduled or not with the variable $x_{ij}(t)$ at time $t$. The scheduling decision action is binary variables (i.e., $x_{ij}(t)=\{0,1\}$) and each scheduling decision result can be formulated as follows,
\begin{equation}
    \mathcal{A}(t) = \{[x(t)]\}.
\end{equation}

\subsubsection{Transition Probability} The transition probability function is formulated as following~\eqref{eq:trans probabiltiy} where the function means that our proposed agent will be converted into the next state $s(t+1)$ when taking scheduling decision action $x_{ij}(t)$ from the current state $s(t)$ with the probability of~\eqref{eq:trans probabiltiy}, i.e.,
\begin{equation}
    P(s(t+1)\mid s(t), a(t)). \label{eq:trans probabiltiy}
\end{equation}

\subsubsection{Reward Function} 
Our considering reward function is equivalent to \eqref{eq:objR} which is fundamentally formulated using \eqref{eq: reward}. 
The reward function is designed for the maximization of the utility/value in \eqref{eq:overall object}, so that our proposed MDP-based scheduling agent determines the discrete-time sequential optimal scheduling decision action $x_{ij}(t)$ for maximizing the reward function value, i.e.,
\begin{eqnarray}
    r(s(t), a(t)) &=& \mathcal{R}(t) \\ 
    &=& \epsilon\cdot R^{T}_{data}(t) - (1-\epsilon)\cdot R^{U}_{power}(t)
\end{eqnarray}
and more details about $R^{T}_{data}(t)$ (reward in towers) and $R^{U}_{power}(t)$ (reward in UAVs) are as follows.

\begin{itemize}
    \item $R^{T}_{data}(t)$: This can be calculated by~\eqref{eq: reward of tower data}. The larger $R^T_{data}(t)$ means that the towers are with more fair data distribution among them. The fair data distribution will ensure that all towers in the system are able to achieve similar and high-performance levels for autonomous object detection deep learning tasks. The data size that each tower has can be obtained by~\eqref{eq: tower data calculation} at time $t$. It is affected by the MDP-based scheduling decision action $x_{ij}(t)$ which is accumulating the transmitted data size from the associated/scheduled UAV to the amount of data where the tower had before (i.e., $d_i(t-1)$). From the tower side of view, $R^T_{data}(t)$ acts as a positive reward.
    \item $R^{U}(t)$: This can be calculated by~\eqref{eq: reward of UAV power}. The reward from the UAV point of view represents the power status for all UAVs compare to the maximum power capacity of UAVs.
    The current power state is also affected by the MDP-based scheduling decision action $x_{ij}(t)$ and the distance between current position and the next position. Here, $j$-th UAV gains additional power from the associated/scheduled $i$-th tower as much as power calculated by~\eqref{eq:wireless_charging_amount} when the MDP-based scheduling decision action $x_{ij}(t)$ is $1$. 
    As a result of the scheduling decision action $x_{ij}(t)$, the larger value of the summation of the power of all UAVs existing in the system $R^U_{power}$ leads to the greater reward values, in the UAV perspective.
 \end{itemize}

\subsubsection{Value Function} The object of the proposed MDP-based workload-aware scheduling algorithm is designed to achieve optimal content delivery decisions between UAVs and towers. We define $\pi:S\rightarrow A$, which maps the current states with a series of actions (e.g., $a = \pi(s)$). For any initial state $s$ and the corresponding policy $\pi \in \Pi$ where $\Pi$ is defined as a set of all stationary policies, the cumulative reward during $\mathcal{T}$ time-step is formulated as follows,
\begin{equation}
    \max_{\pi \in \Pi}:\sum_{t=1}^{\mathcal{T}}\gamma^{t}r(s^\pi(t), a(t))
\end{equation}
where $0\leq \gamma \leq 1$. Based on the transition probability and cumulative reward, the value function $V$ is defined as, 
\begin{equation}
    V^*(s) = \max_{a\in A}\{r(s,a) + \gamma^t\sum_{s'\in S}P(s'\mid s,a)V^*(s')\}. \label{eq:value function}
\end{equation}

\begin{figure}
    \centering
    \includegraphics[width=0.8\linewidth]{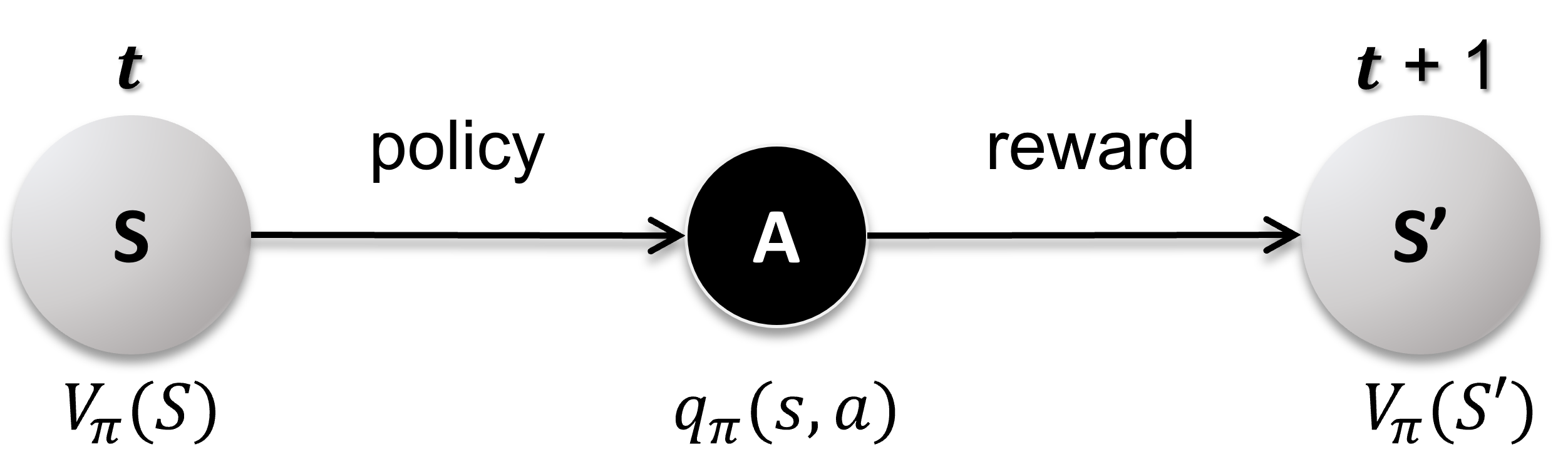}
    \caption{The value function of MDP.}
    \label{fig:mdp}
\end{figure}
%and the process is presented in Algorithm~\ref{algo:MDP_scheduling}.

% \begin{algorithm}[t!]
% \small
% %\setstretch{1.5}
% \caption{AoI aware Scheduling for Content Access}
% \label{algo:MDP_scheduling}
%  \textbf{Input:} reward function $r(s(t), a(t))$, transitional model $P(s'|s,a)$, discounted factor $\gamma$, convergence threshold $\theta$\\
%  \textbf{Output:}optimal policy $\pi^{*}$\\
%  Initialize $V(s)$ with zeros\\
%  Converge $\leftarrow$ false
%  \While{converge $=$ false}{
%     \STATE $\Delta \leftarrow 0$
%     \For{$s \in S$}{
%         temp $\leftarrow v(s)$
%         $v(s) \leftarrow r(s,a) + \gamma^t\sum_{s'\in S}P(s'\mid s,a)V^*(s')$
%         $\Delta \leftarrow \max(\Delta, |temp-v(s)|)$
%         }
%     \If{$\Delta <\theta$}{
%         converge $\leftarrow$ true
%         }
%     }
%  \For{$s\in S$}{
%     $\pi^{*}(s) \leftarrow argmax\sum_{s'\in S}P(s'\mid s,a)V^*(s')$
%     }
%  \textbf{Return} $\pi^{*}$
% \end{algorithm}

\section{Performance Evaluation}\label{sec:4}
This section describes our simulation setup for evaluating the proposed workload-aware MDP-based scheduling algorithm and its performance evaluation results.

\begin{figure*}[t]\centering
    \begin{multicols}{7}
     \includegraphics[width=0.995\linewidth]{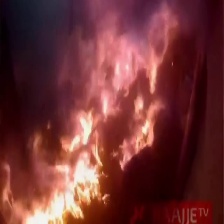}\captionsetup{justification=centering}
     \subcaption{Fire 1}
     \includegraphics[width=0.995\linewidth]{{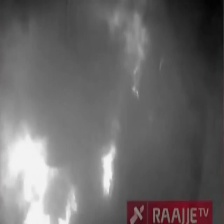}}\captionsetup{justification=centering}
     \subcaption{Fire 2}
     \includegraphics[width=0.995\linewidth]{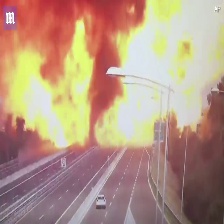}\captionsetup{justification=centering}
     \subcaption{Fire 3}
     \includegraphics[width=0.995\linewidth]{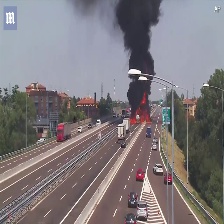}\captionsetup{justification=centering}
     \subcaption{Fire 4}
     \includegraphics[width=0.995\linewidth]{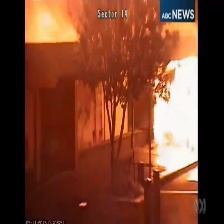}\captionsetup{justification=centering}
     \subcaption{Fire 5}
     \includegraphics[width=0.995\linewidth]{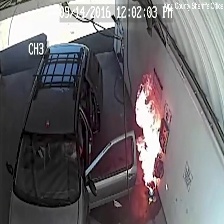}\captionsetup{justification=centering}
     \subcaption{Fire 6}
     \includegraphics[width=0.995\linewidth]{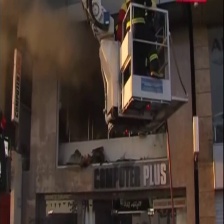}\captionsetup{justification=centering}
     \subcaption{Fire 7}
     \end{multicols}
     
     \begin{multicols}{7}
     \includegraphics[width=0.995\linewidth]{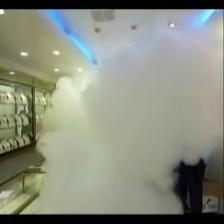}\captionsetup{justification=centering}
     \subcaption{Smoke 1}
     \includegraphics[width=0.995\linewidth]{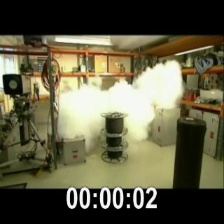}\captionsetup{justification=centering}
     \subcaption{Smoke 2}
     \includegraphics[width=0.995\linewidth]{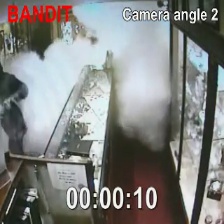}\captionsetup{justification=centering}
     \subcaption{Smoke 3}
     \includegraphics[width=0.995\linewidth]{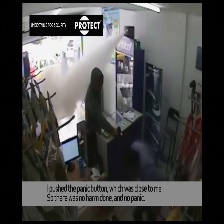}\captionsetup{justification=centering}
     \subcaption{Smoke 4}
     \includegraphics[width=0.995\linewidth]{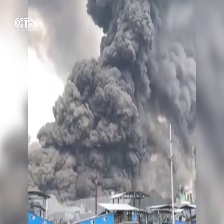}\captionsetup{justification=centering}
     \subcaption{Smoke 5}
     \includegraphics[width=0.995\linewidth]{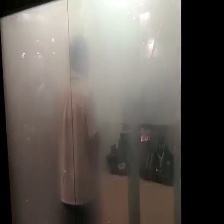}\captionsetup{justification=centering}
     \subcaption{Smoke 6}
     \includegraphics[width=0.995\linewidth]{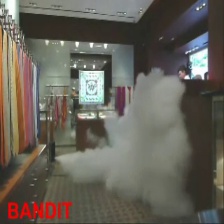}\captionsetup{justification=centering}
     \subcaption{Smoke 7}
     \end{multicols}

     \begin{multicols}{7}
     \includegraphics[width=0.995\linewidth]{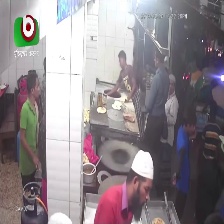}\captionsetup{justification=centering}
     \subcaption{Default 1}
     \includegraphics[width=0.995\linewidth]{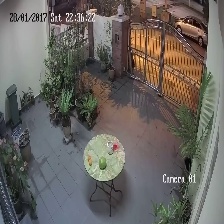}\captionsetup{justification=centering}
     \subcaption{Default 2}
     \includegraphics[width=0.995\linewidth]{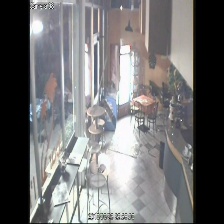}\captionsetup{justification=centering}
     \subcaption{Default 3}
     \includegraphics[width=0.995\linewidth]{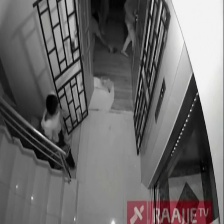}\captionsetup{justification=centering}
     \subcaption{Default 4}
     \includegraphics[width=0.995\linewidth]{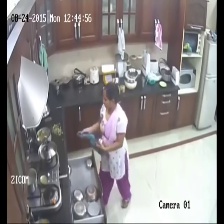}\captionsetup{justification=centering}
     \subcaption{Default 5}
     \includegraphics[width=0.995\linewidth]{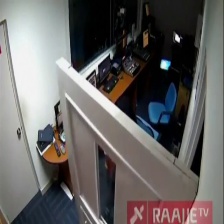}\captionsetup{justification=centering}
     \subcaption{Default 6}
     \includegraphics[width=0.995\linewidth]{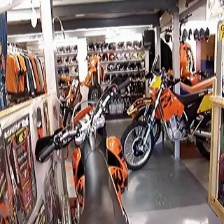}\captionsetup{justification=centering}
     \subcaption{Default 7}
     \end{multicols}
     
\caption{Examples of training images in three classes; fire, smoke, and default, respectively. Our proposed algorithm helps the training models to differentiate between flame cases (fire class) and non-flame cases (smoke and default classes) more precisely~\cite{dataset}.}
\label{fig:training images}
\end{figure*}

\subsection{Simulation Setup}\label{sec:4-1}

\begin{table}[t]
\footnotesize
\caption{Specification of simulation platforms.}
\label{tab:param}
\begin{center}
	\centering
	\begin{tabular}{l|l}
    \toprule[1.0pt]
    \centering
      System  & Specification \\
    \midrule[1.0pt]
    CPU & Intel(R) Core(TM) i7-9700K CPU @3.60 GHz \\
        & RAM: 64.0 GB \\ 
    \midrule
    GPU & NVIDIA GeForce RTX 3090 \\ 
        & The number of cores: $10,496$ \\
        & Memory: 24 GB GDDR6X\\
    \midrule
    Platform (PC)  & CPU: Intel i7-9700K (3.60 GHz) \\
                   & Memory: DDR4 64 GB \\
                   & SSD: 500 GB (NVMe) \\
                   & HDD: 1 TB \\
                   & VGA: RTX 3090 \\
    \bottomrule[1.0pt]
	\end{tabular}
\end{center}
\end{table}

The performance of the proposed workload-aware MDP-based scheduling algorithm is evaluated in two-dimensional grid map environment, as illustrated in Fig.~\ref{fig:period (i)}.
The proposed system consists of $10$ UAVs and $5$ towers. The size of the area is $1250\,m \times 1250\,m$, and there are $100$ regions on the map. The sizes of individual regions are equivalent and they have different amounts of information via randomly occurring damage and abnormal behavior events, as shown in Fig.~\ref{fig:training images}. 
We place the towers evenly over the entire network area, and the intervals between the towers are all equivalently constant. 
The initial positions of $10$ UAVs are determined according to various trajectory models~\cite{electronics, elec[75]}. The UAVs' waypoints over time are summarized in Table~\ref{tab:uav position} at $10$-minute intervals. Each UAV moves between waypoints at flight speeds during the performance evaluation time slots. We consider the UAV models as a DJI Phantom4 Pro v2.0 UAV (DJI, Shenzhen, China)~\cite{elec[66]} whose parameters are specified in Table~\ref{tab:parameters_of_uav}. Each value is applied to the data acquisition and power consumption models of UAVs defined above. 
We also assume that $100$ regions have different situations, and the states can be changed randomly to validate our proposed algorithm applied in general situations.
The UAVs collect information on the region along the pre-determined paths and transmit the generated contents to the associated/scheduled towers. During the entire performance evaluation process, we limit the number of channels to connect with one UAV at a time in the tower. Lastly, our simulation platform setting is presented in Table~\ref{tab:param}.
 
In order to evaluate the performance of the proposed algorithm, we focus on the following indicators as performance evaluation criteria, 
\begin{itemize}
    \item \textit{(Criteria 1)} UAVs' efficient power charging to maintain multi-UAV networks, i.e., \textit{Power-Charging Efficiency at UAVs}.
    \item \textit{(Criteria 2)} The amount and equitable degree of each tower's data collection, i.e., \textit{Data Distribution Fairness at Towers}.
    \item \textit{(Criteria 3)} The accuracy of the autonomous object recognition (for abnormal behavior detection) deep learning models using the data delivered by the proposed scheduling algorithm, i.e., \textit{Learning Accuracy at Towers}.
\end{itemize}

As mentioned above, all towers utilize training models and this paper considers ResNet50~\cite{resnet50} and VGG16~\cite{vgg16} which are the representative convolutional neural network (CNN) object recognition models. Both CNN models classify flame cases and non-flame cases, as shown in Fig.~\ref{fig:training images} using the data collected by multi-UAVs and the corresponding hyper-parameters are described and summarized in Table~\ref{tab:hyper-parameters}. This paper considers that all UAVs collect image data such as Fig.~\ref{fig:training images} by built-in sensors in different regions. 
In general, the accuracy of the learning models is determined by the amount of data used for training as long as there is no overfitting. As seen in previous studies, the accuracy increases according to the input data size and converges to a certain level of performance.
Through this, when the towers acquire several regions' data from UAVs according to the proposed scheduling algorithm and train the model, it is anticipated that all towers will have similar learning performance. 
In addition, it is expected that all towers will take as many data sizes as possible during the resource exchange period, as UAVs are generally operated through sufficiently charged power from the tower.  Furthermore, our proposed scheduling algorithm helps the multi-UAVs collect data evenly over time based on the fundamental objectives as formulated in the optimization framework. 

To validate that our proposed algorithm efficiently helps the model train itself by controlling UAVs, we benchmark our proposed scheduling algorithm with three comparison algorithms as follows.
\begin{itemize}
    \item \textit{Proposed Algorithm:} As described in Sec.~\ref{sec:3-2}, all UAVs equally consider both two values, $R^{T}_{data}(t)$ (rewards in tower side) and $R^{U}_{power}(t)$ (rewards in UAV side) with $\epsilon=0.5$. In other words, all UAVs try to transmit their collected data to towers evenly while taking into account the remaining energy.
    \item \textit{Comp1 Algorithm:} The Comp1 algorithm considers only $R^U_{power}(t)$ using~\eqref{eq: reward of UAV power} as the reward function where the value of $\epsilon$ is 0. Therefore, all UAVs concentrate on saving their own remaining energy.
    \item \textit{Comp2 Algorithm:} The Comp2 algorithm considers only $R^T_{data}(t)$ using~\eqref{eq: reward of tower data} as the reward function where the value of $\epsilon$ is 1 opposite to the Comp1 algorithm. Therefore, all UAVs scheduled by the Comp2 algorithm try to transmit their collected data evenly to towers as much as possible.
    \item \textit{Comp3 Algorithm:} All UAVs transmit its collected data to towers randomly, i.e., similar to random walk computation. In other words, this random algorithm performs multi-UAV and multi-tower scheduling randomly without any other considerations. 
\end{itemize}

\begin{table*}[t]
%\small
\centering
\caption{Specification of UAV model \cite{elec[75]}. }
\renewcommand{\arraystretch}{1.0}
\begin{tabular}{c||c}
\toprule[1pt]
\textbf{Notation} & \textbf{Value} \\ \midrule
Flight speed, $v$ & 20 m/s \\
Capacity of flight battery & 5,870 mAh\\
Voltage & 17.4 V\\
Aircraft weight with battery and propellers, $W$ & 1,375 g\\ Rotor radius, $R$ & 0.4 m \\
Rotor disc area, $A=\pi R^{2}$ & 0.503 $m^{2}$ \\
Number of blades , $b$ & 4 \\
Rotor solidity, $s, \frac{0.0157b}{\pi R}$ & 0.05 \\
Blade angular velocity, $\Omega$ & 300 radius/s\\
Tip speed of the rotor blade , $U_{tip}=\Omega R^{2}$ & 120 \\
Fuselage drag ratio, $d_{0}=\frac{0.0151}{sA}$ & 0.6 \\
Air density, $\rho$ & 1.225 kg/$m^{3}$ \\
Mean rotor-induced velocity in hovering, $v_{0}=\sqrt\frac{W}{s\rho A}$ & 4.03 \\
Profile drag coefficient, $\delta$ & 0.012 \\
Incremental correction factor to induced power, $k$ & 0.1 \\
\bottomrule[1pt]
\end{tabular}
\label{tab:parameters_of_uav}
\end{table*}

\begin{table}[t]
%\small
\centering
\caption{Specification of training hyper-parameters.}
\renewcommand{\arraystretch}{1.0}
\begin{tabular}{c||c}
\toprule[1pt]
\textbf{Notation} & \textbf{Value} \\ \midrule
Model & ResNet50, VGG16 \\
Optimizer & Adam \\
Loss function & Cross-entropy \\
Activation function & Softmax \\
Batch size & 77 \\
Number of epochs & 20 \\
Number of classes & 3 \\
\bottomrule[1pt]
\end{tabular}
\label{tab:hyper-parameters}
\end{table}

\begin{table*}[t]
\scriptsize
\caption{Waypoints of UAVs~\cite{electronics}.}
    \centering
        \begin{tabular}{c|cccccccccc}
        \toprule[1pt]
        Time (min) & UAV \#1 & UAV \#2 & UAV \#3 & UAV \#4 & UAV \#5 & UAV \#6 & UAV \#7 & UAV \#8 & UAV \#9 & UAV \#10 \\ \midrule
        10 & (125, 1075) & (375, 1175) & (275, 800) & (625, 475) & (100, 100) & (625, 225) & (550, 1075) & (475, 800) & (150, 825) & (650, 1200) \\
        20 & (625, 1075) & (200, 1075) & (175, 625) & (575, 250) & (150, 200) & (125, 225) & (625, 900) & (575, 625) & (175, 1025) & (600, 1100) \\
        30 & (625, 650) & (125, 900) & (275, 425) & (375, 150) & (200, 300) & (125, 625) & (625, 650) & (475, 425) & (375, 1100) & (550, 1100) \\
        40 & (120, 650) & (125, 650) & (475, 425) & (175, 250) & (250, 400) & (625, 625) & (625, 400) & (275, 425) & (575, 1025) & (500, 900) \\
        50 & (120, 225) & (125, 400) & (575, 625) & (150, 475) & (300, 500) & (625, 1075) & (550, 225) & (175, 625) & (625, 825) & (450, 800) \\
        60 & (625, 225) & (200, 225) & (475, 800) & (375, 650) & (350, 600) & (125, 1075) & (375, 150) & (275, 800) & (375, 650) & (400, 700) \\
        70 & (125, 225) & (375, 150) & (275, 800) & (625, 825) & (400, 700) & (625, 1075) & (200, 225) & (475, 800) & (150, 475) & (350, 600) \\
        80 & (125, 650) & (550, 225) & (175, 625) & (575, 1025) & (450, 800) & (625, 650) & (125, 400) & (575, 625) & (175, 250) & (300, 500) \\
        90 & (625, 650) & (625, 400) & (275, 425) & (375, 1100) & (500, 900) & (125, 650) & (125, 650) & (475, 425) & (375, 150) & (250, 400) \\
        100 & (625, 1075) & (625, 650) & (475, 425) & (175, 1025) & (550, 1000) & (125, 225) & (125, 900) & (275, 425) & (575, 250) & (200, 300) \\
        % 110 & (125, 1075) & (625, 900) & (575, 625) & (150, 825) & (600, 1100) & (625, 225) & (200, 1075) & (175, 625) & (625, 475) & (150, 200) \\
        % 120 & (625, 1075) & (550, 1075) & (475, 800) & (625, 475) & (650, 1200) & (125, 225) & (375, 1175) & (275, 800) & (625, 475) & (100, 100) \\ 
        \bottomrule[1pt]
        \end{tabular}
    \label{tab:uav position}
\end{table*}

\subsection{Simulation Results}\label{sec:4-2}
 
This section describes various performance results of our proposed workload-aware MDP-based scheduling decision algorithm, in terms of power-charging efficiency at UAVs (refer to Sec.~\ref{sec:4-2-1}), data distribution fairness at towers (refer to Sec.~\ref{sec:4-2-2}), and learning accuracy at towers (refer to Sec.~\ref{sec:4-2-3}), respectively. 

\subsubsection{Power-Charging Efficiency at UAVs}\label{sec:4-2-1}
\begin{figure}
    \centering
    \includegraphics[width=0.9\linewidth]{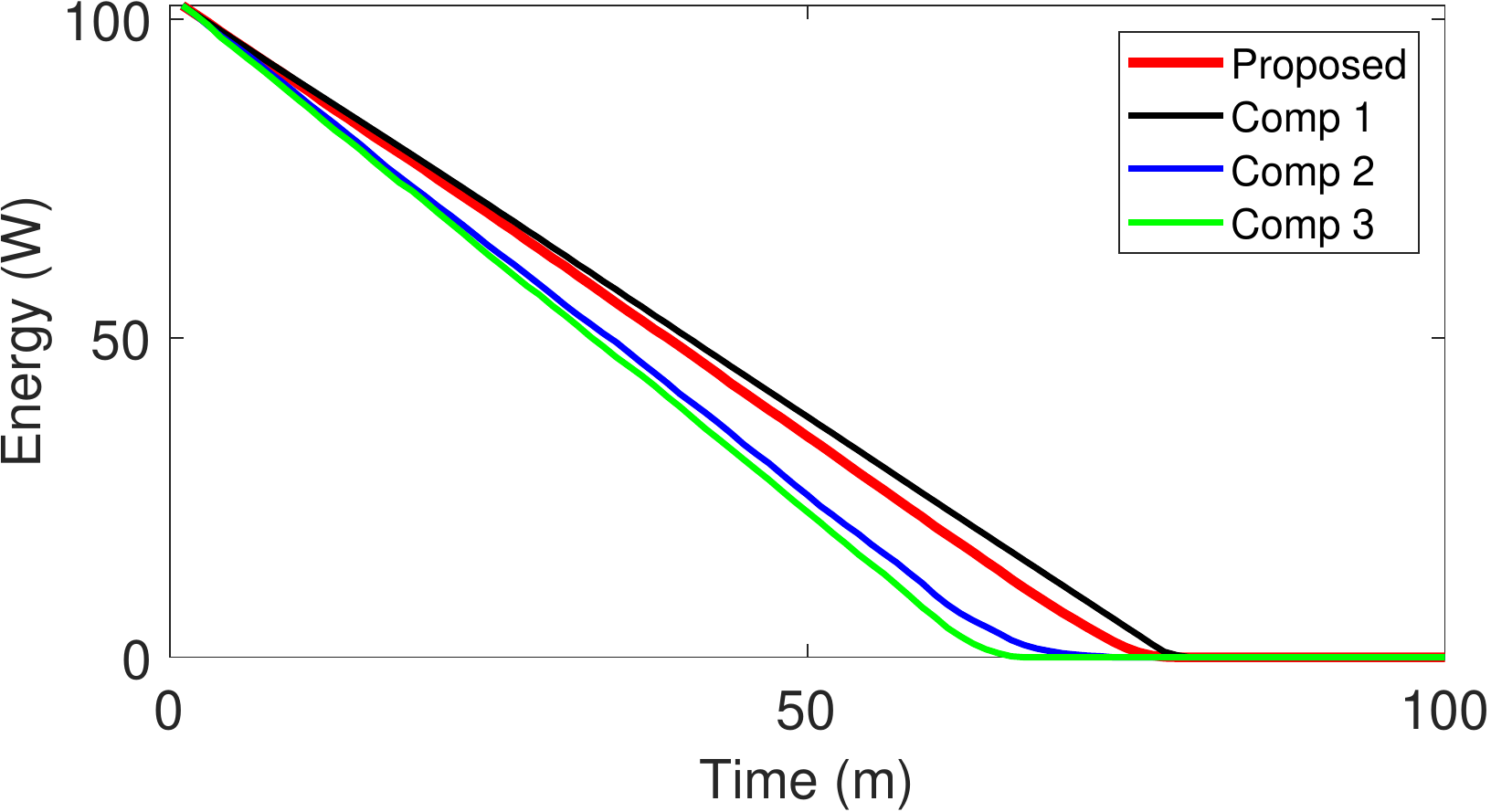}
    \caption{Average remaining energy of all UAVs in each scheduling method over time.}
    \label{fig:UAV Energy}
\end{figure}

% Energy Consumption
Fig.~\ref{fig:UAV Energy} shows the average power decrease of all $10$ UAVs in multi-UAV networks. First of all, we show that the Comp3 algorithm has the worst maintenance because each UAV's energy status are not considered at all. The Comp2 algorithm does not treat securing power as essential and tends to be similar to the worst case. In contrast, the Comp1 algorithm maximizes the UAVs' remaining power over time. As a result, the operations of UAV-based networks are most reliably guaranteed and preserved during the period of resource exchange. The proposed algorithm, which focuses on the two values equally with $\epsilon=0.5$, shows the performance close to the case where the system reliability is maintained during the longest time.

\subsubsection{Data Distribution Fairness at Towers}\label{sec:4-2-2}
\begin{figure}
    \centering
    \includegraphics[width=\linewidth]{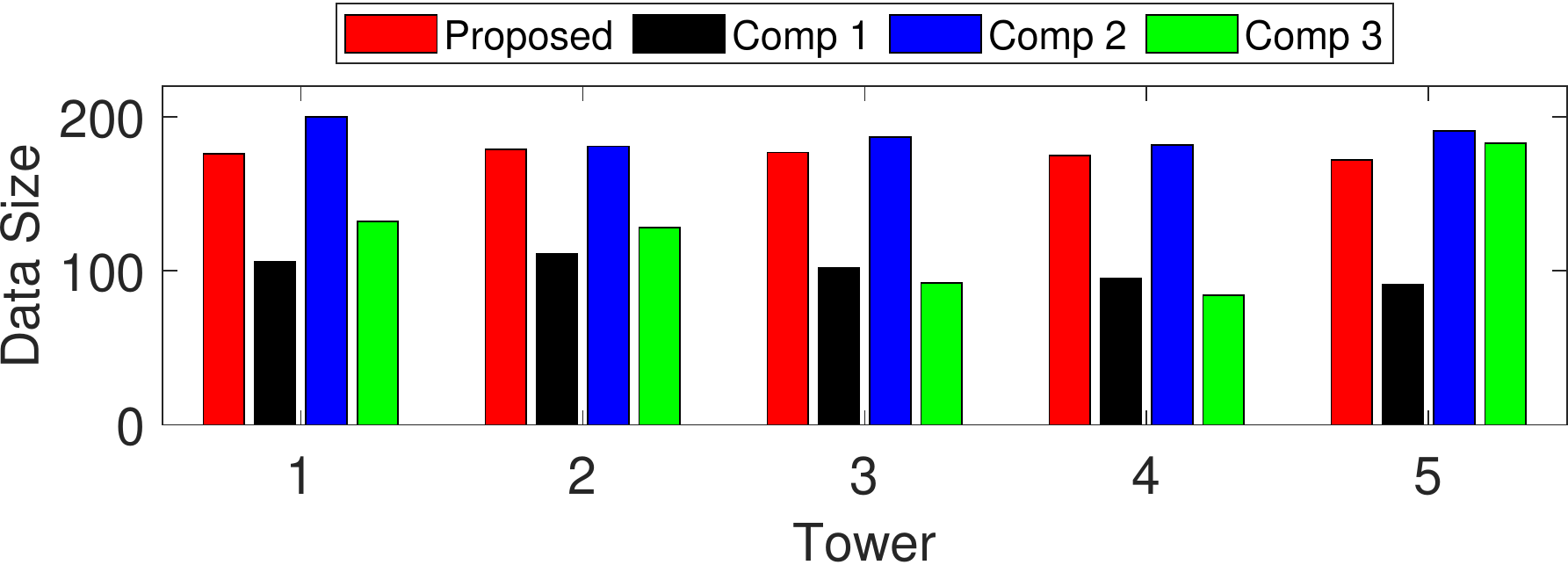}
    \caption{Total data size of all towers in each scheduling algorithm.}
    \label{fig:Total Data Size}
\end{figure}

\begin{figure}[t]\centering
\centering
    \begin{multicols}{2}
     \includegraphics[width=\linewidth]{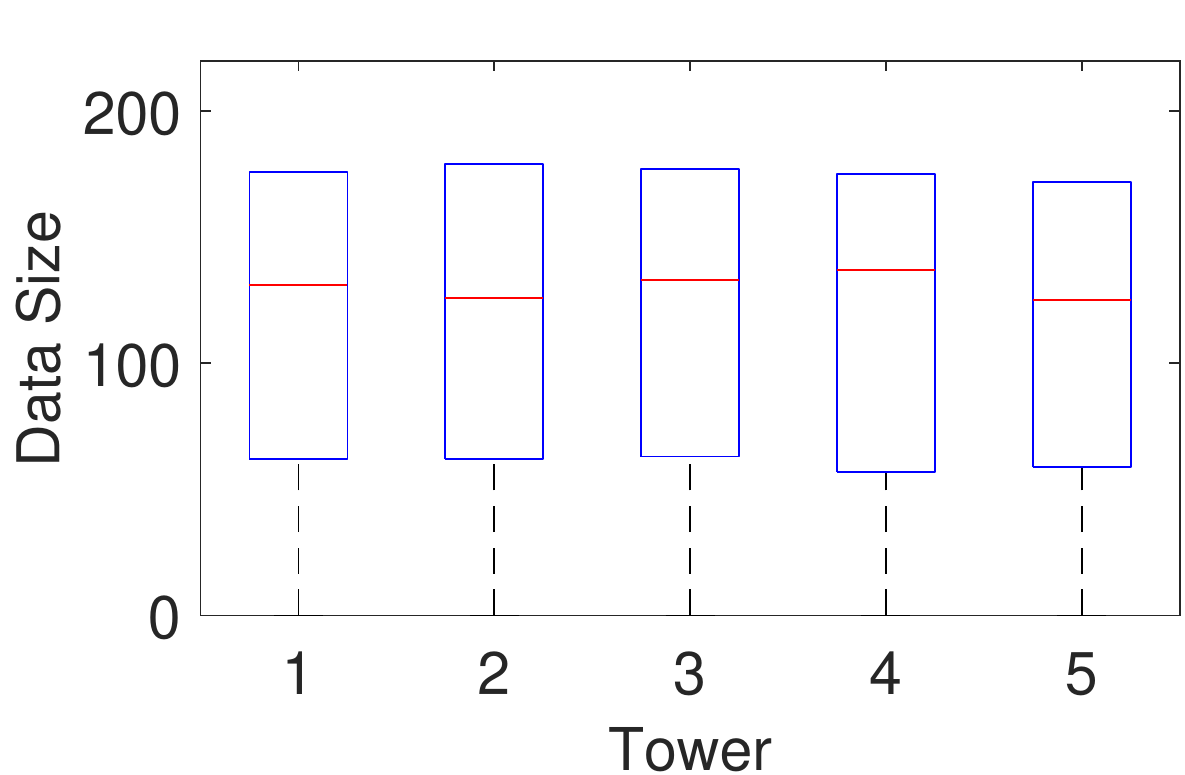}\captionsetup{justification=centering}
     \subcaption{\textbf{Proposed}}
     \includegraphics[width=\linewidth]{{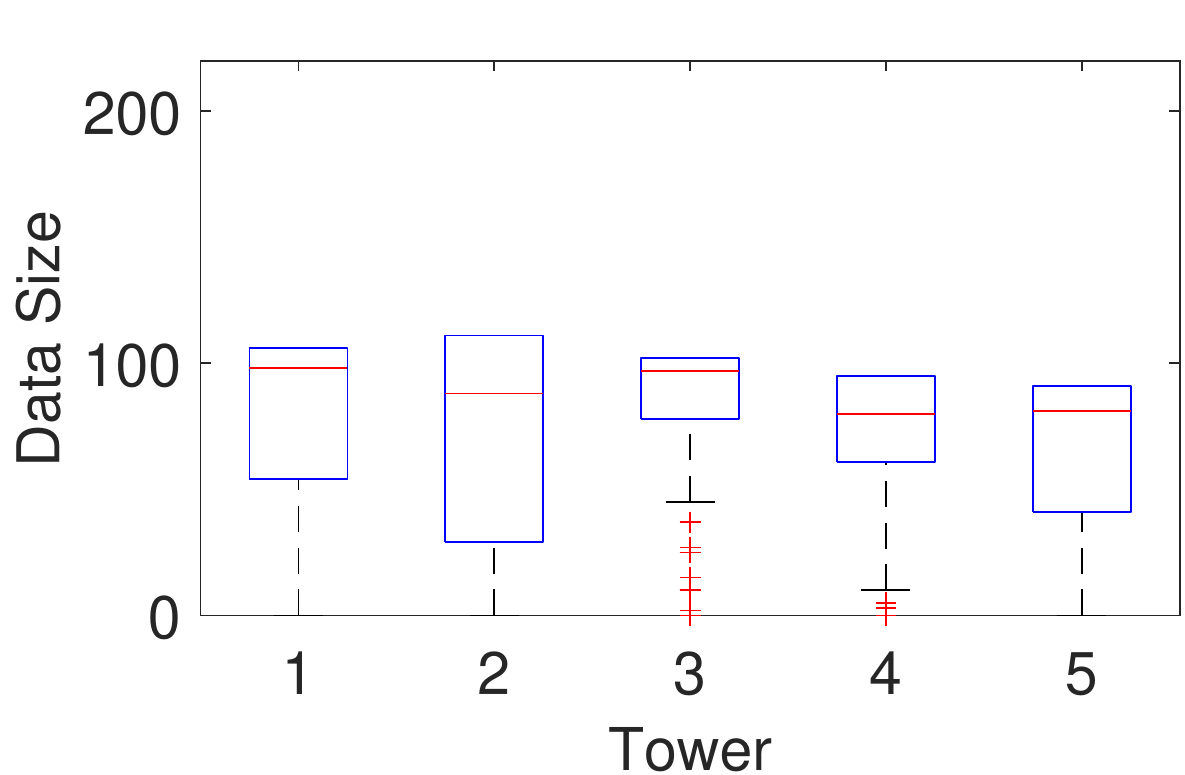}}\captionsetup{justification=centering}
     \subcaption{Comp 1}
    \end{multicols}
    \begin{multicols}{2}
    \includegraphics[width=\linewidth]{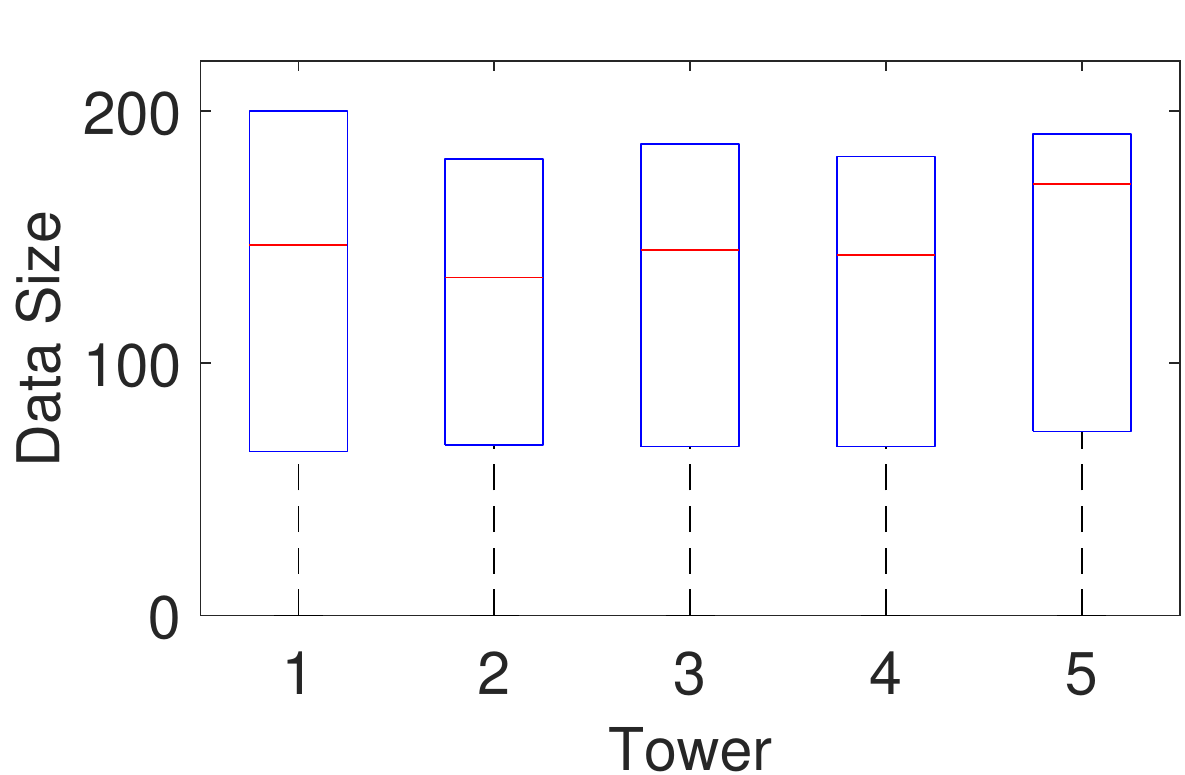}\captionsetup{justification=centering}
     \subcaption{Comp2}
     \includegraphics[width=\linewidth]{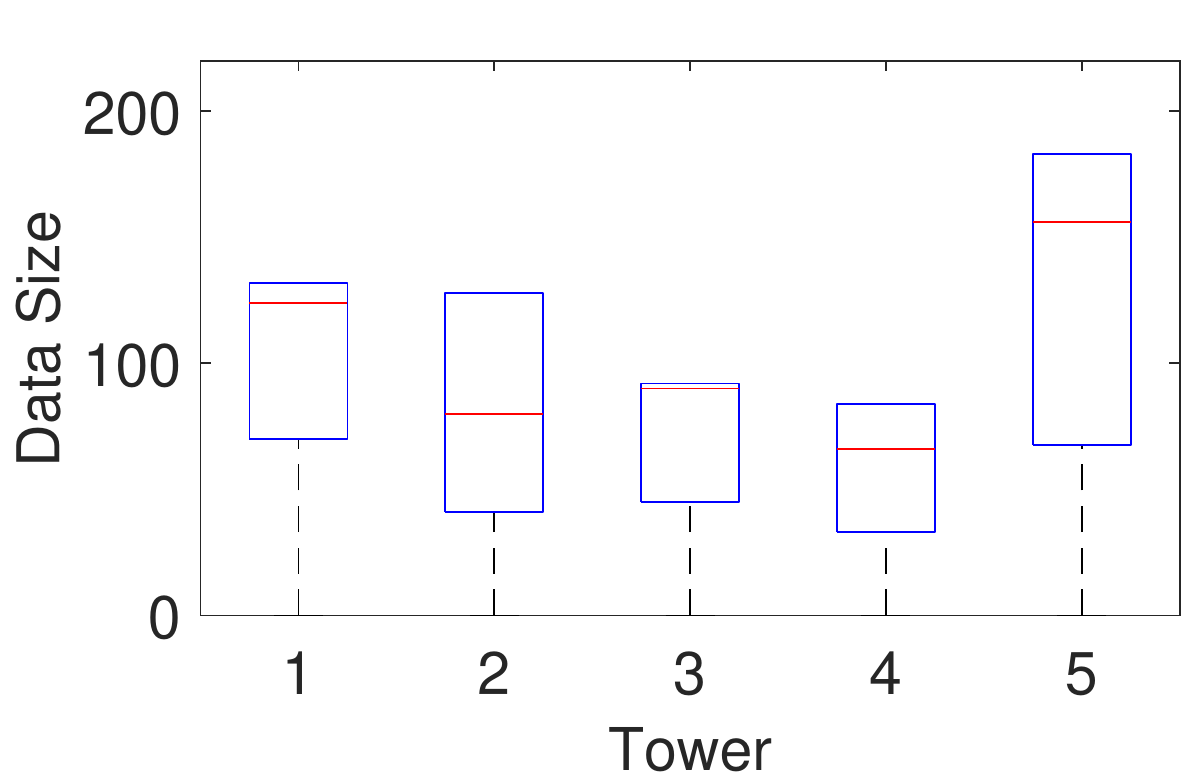}\captionsetup{justification=centering}
     \subcaption{Comp3}
    \end{multicols}
\caption{Total data size of all towers over operation time.}%
\label{fig:Box Plot}
\end{figure}

% Data size
The performance in terms of the amount of training data that the towers have can be confirmed through Figs.~\ref{fig:Total Data Size}--\ref{fig:Box Plot}.
Each figure shows all towers' accumulated data size obtained by UAVs which are scheduled by various algorithms during $60$ minutes.
In Fig.~\ref{fig:Total Data Size}, all towers in the proposed scheduling algorithms with different $\epsilon$ values except the random algorithm tend to occupy data size evenly. Especially, the proposed and Comp2 algorithms ensure that towers collect as much data as possible. This is because the above two scheduling algorithms only consider even data distributions among all towers, as formulated in~\eqref{eq: reward of tower data}. Here, all towers in the Comp2 algorithm have larger data slightly than the proposed algorithm because the proposed algorithm simultaneously considers the power/energy consumption of all UAVs.

% Box Plot
In Fig.~\ref{fig:Box Plot}, we can show the maximum, minimum, and average values of all towers' data collections by UAVs in each scheduling algorithm. Similar to the results of Fig.~\ref{fig:Total Data Size}, Figs.~\ref{fig:Box Plot}(b)--(c) show that a large amount of data is fairly distributed. In the other two cases, the average data secured in the other two cases is small, and the deviation between towers is more significant than that of other algorithms.
The UAVs scheduled by the proposed algorithm collect data more evenly over operation times than the Comp2 algorithm because the overall red lines representing the median value in the proposed algorithm are more central than the Comp2 algorithm. It means that all UAVs scheduled by the proposed algorithm transmit collected data to their associated/scheduled towers more consistently than the Comp2 algorithm over time. In order to consider overall high-performance surveillance situations, UAVs must transmit data evenly for all operation times.
Accordingly, the proposed algorithm always presents performance approximating the best case regarding UAV power efficiency and tower data collection. Our proposed workload-aware MDP-based scheduling decision algorithm considers the two aforementioned aspects and guarantees the most optimal scheduling decision actions over time.

\begin{figure*}[t]\centering
    \includegraphics[width=0.5\linewidth]{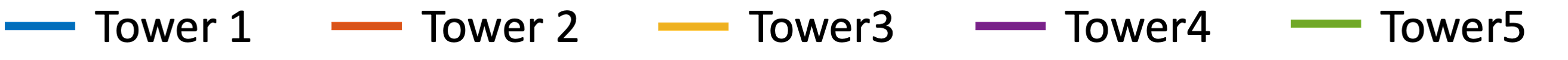}\captionsetup{justification=centering}\\
    \begin{multicols}{2}
     \includegraphics[width=0.8\linewidth]{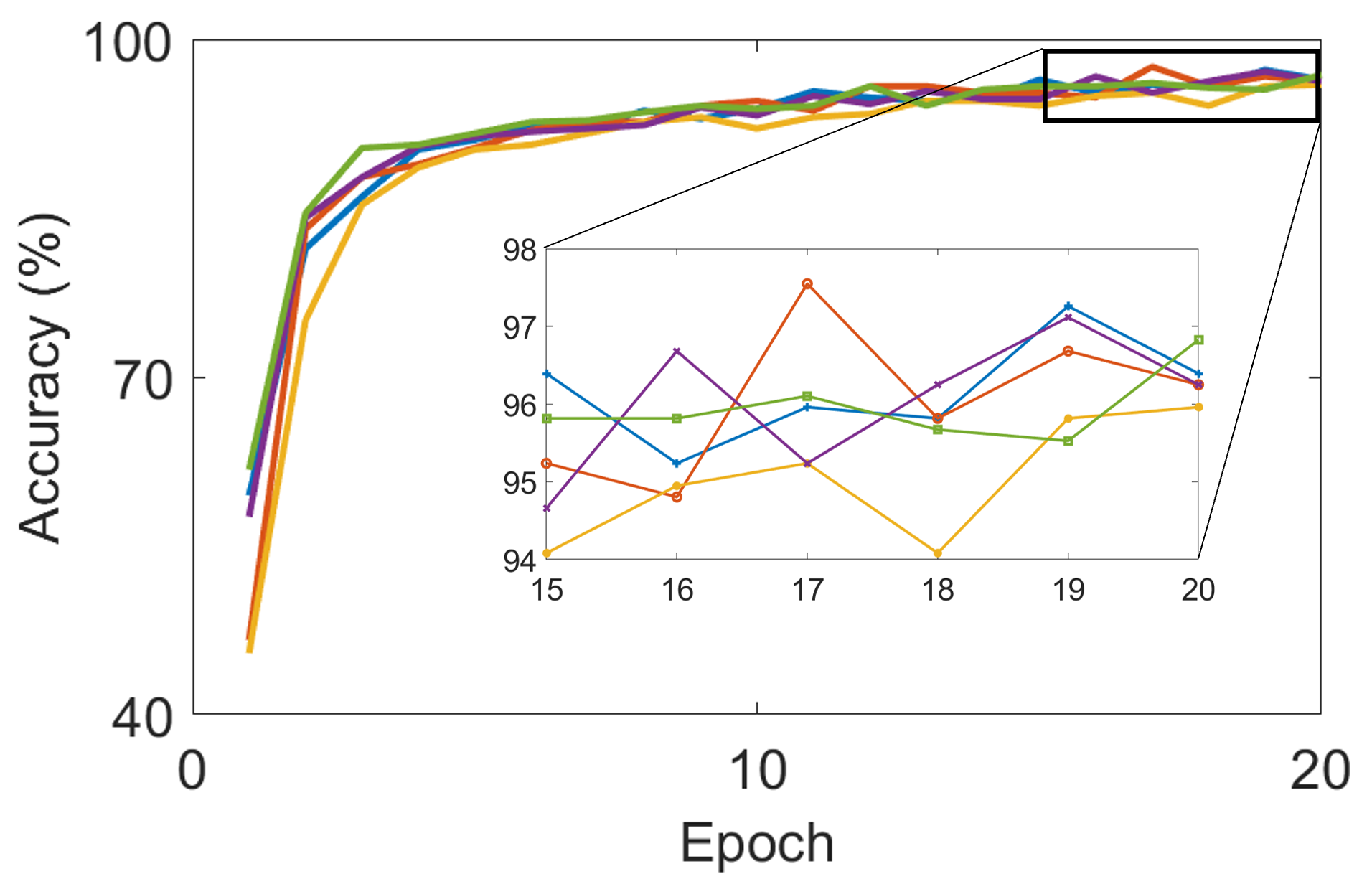}\captionsetup{justification=centering}
     \subcaption{\textbf{Proposed}}
     \includegraphics[width=0.8\linewidth]{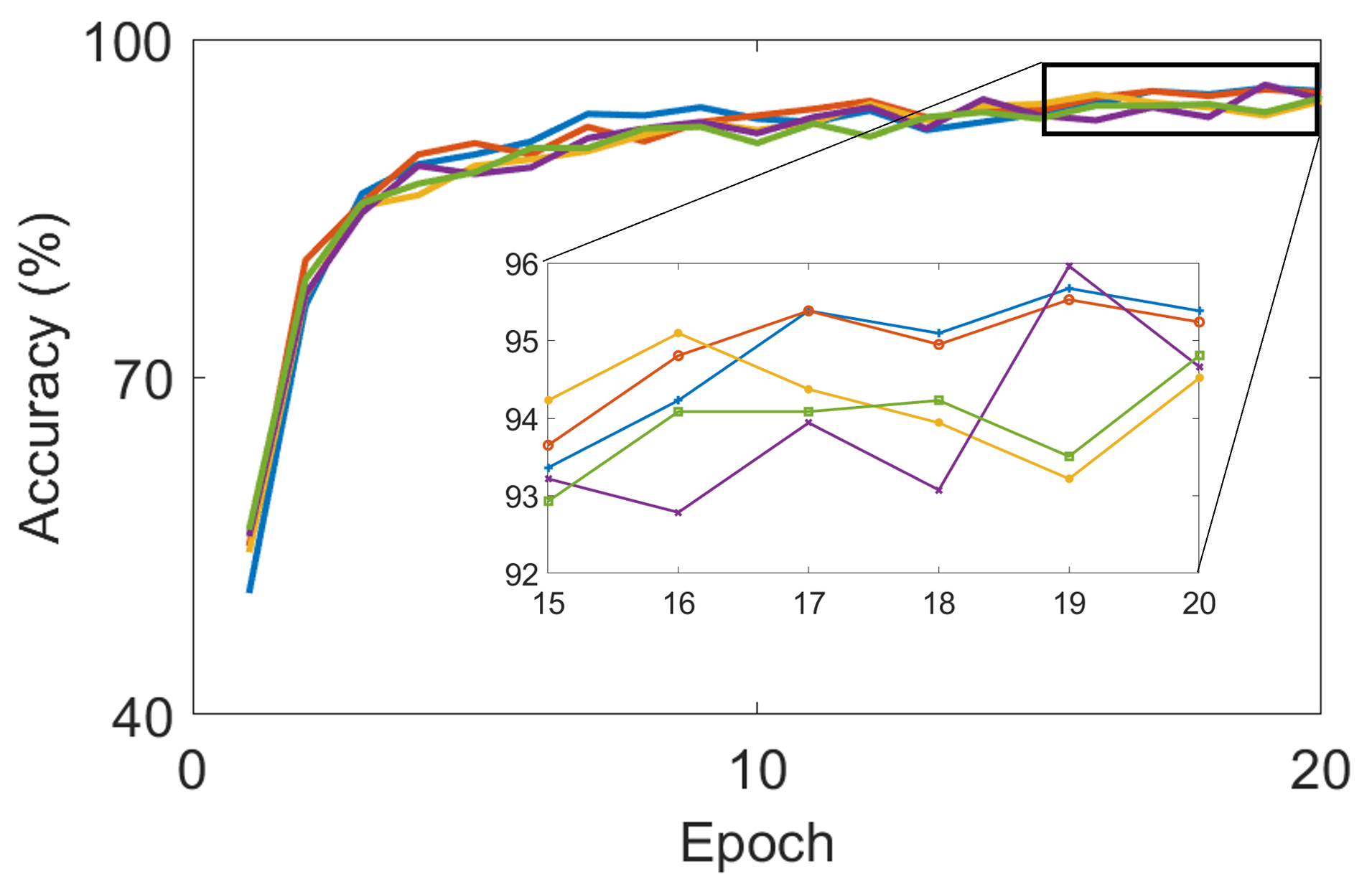}\captionsetup{justification=centering}
     \subcaption{Comp 1}
     \end{multicols}
     \begin{multicols}{2}
    \includegraphics[width=0.8\linewidth]{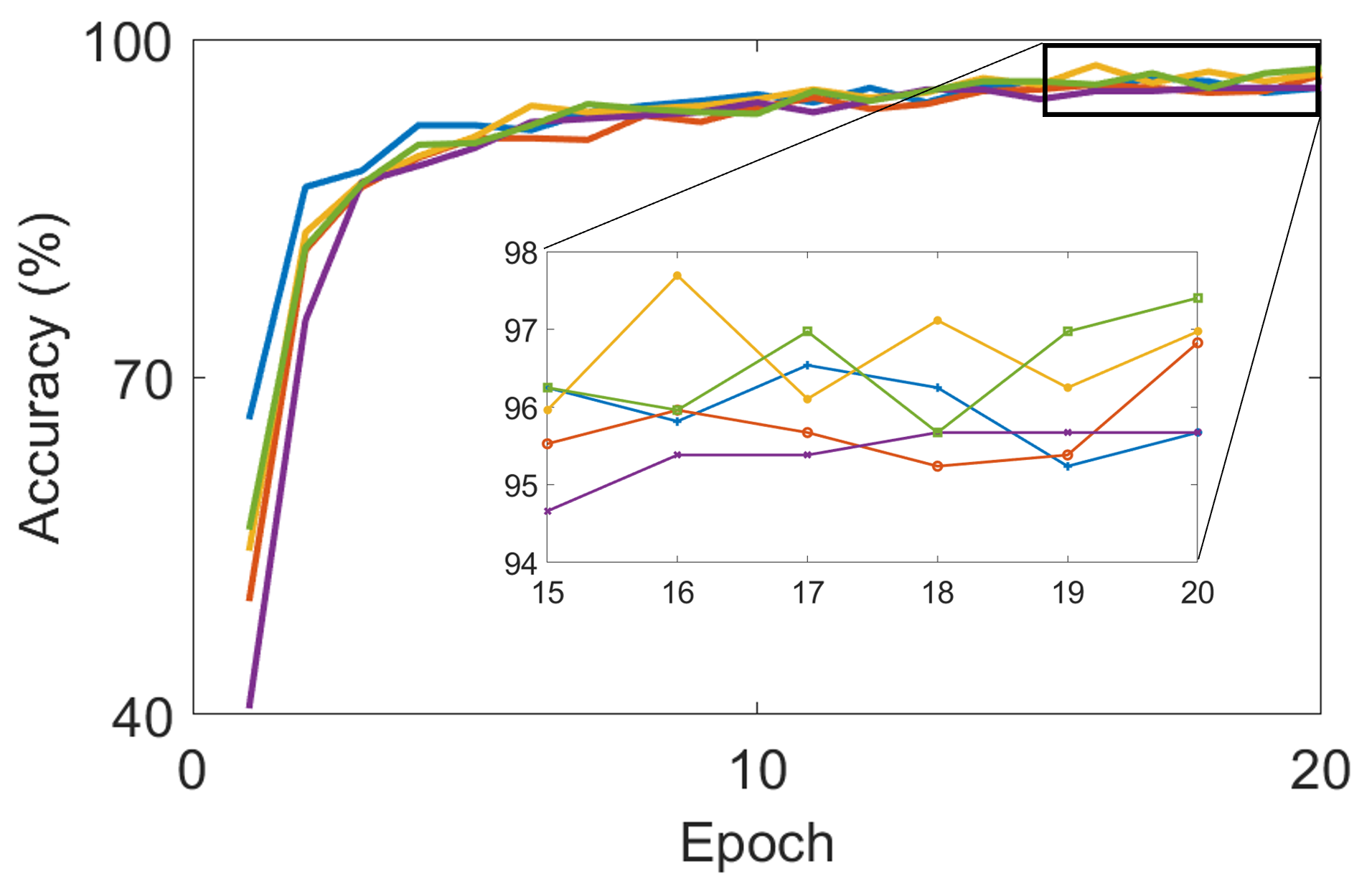}\captionsetup{justification=centering}
     \subcaption{Comp 2}
     \includegraphics[width=0.8\linewidth]{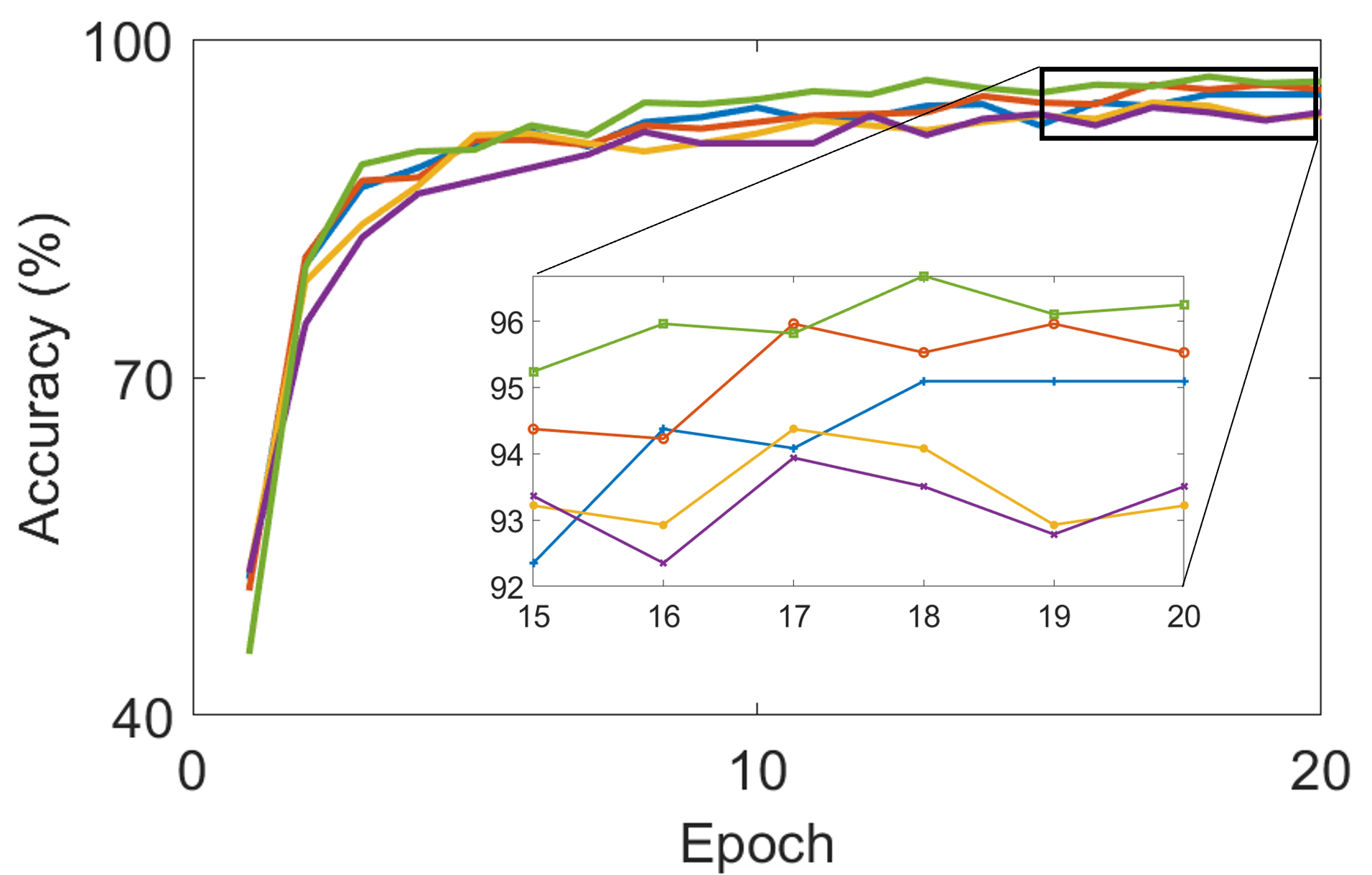}\captionsetup{justification=centering}
     \subcaption{Comp 3}
    \end{multicols}
\caption{Training results of each scheduling algorithm in ResNet50.}%
\label{fig:training_result_in_ResNet50}
\end{figure*}

\begin{figure*}[t]\centering
    \includegraphics[width=0.5\linewidth]{figure/Index_Tower.png}\captionsetup{justification=centering}\\
    \begin{multicols}{2}
     \includegraphics[width=0.8\linewidth]{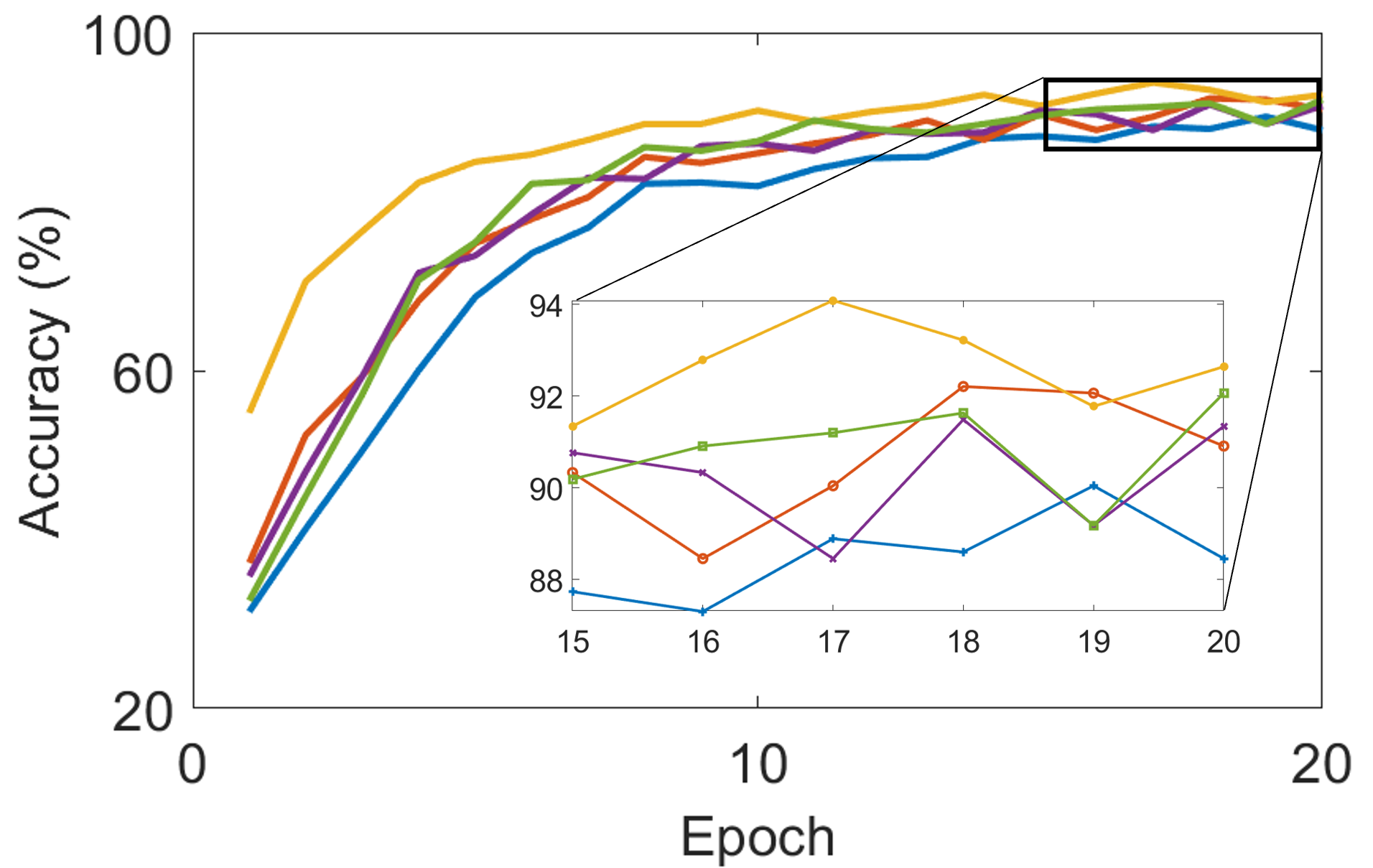}\captionsetup{justification=centering}
     \subcaption{\textbf{Proposed}}
     \includegraphics[width=0.8\linewidth]{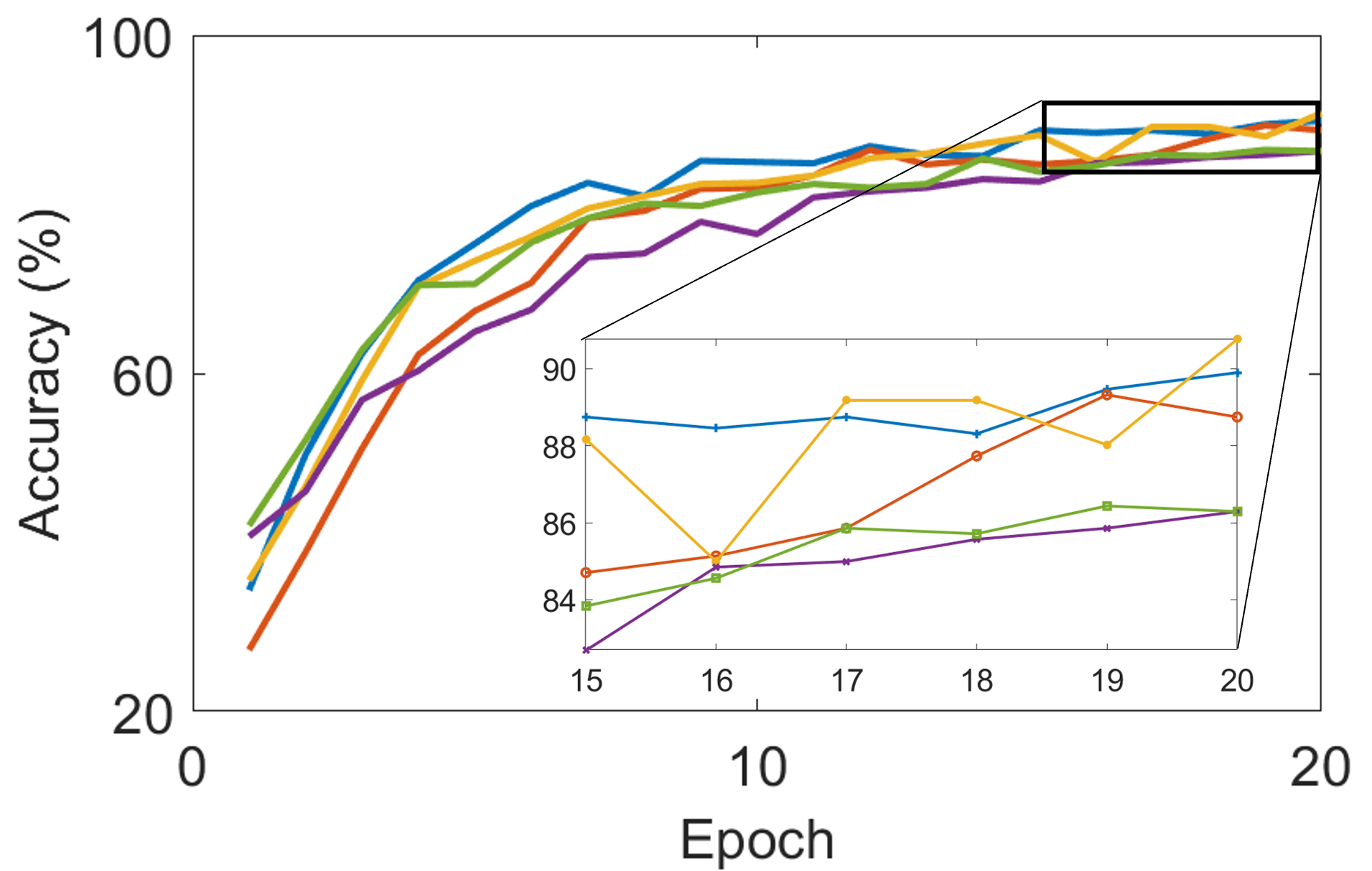}\captionsetup{justification=centering}
     \subcaption{Comp 1}
     \end{multicols}
     \begin{multicols}{2}
    \includegraphics[width=0.8\linewidth]{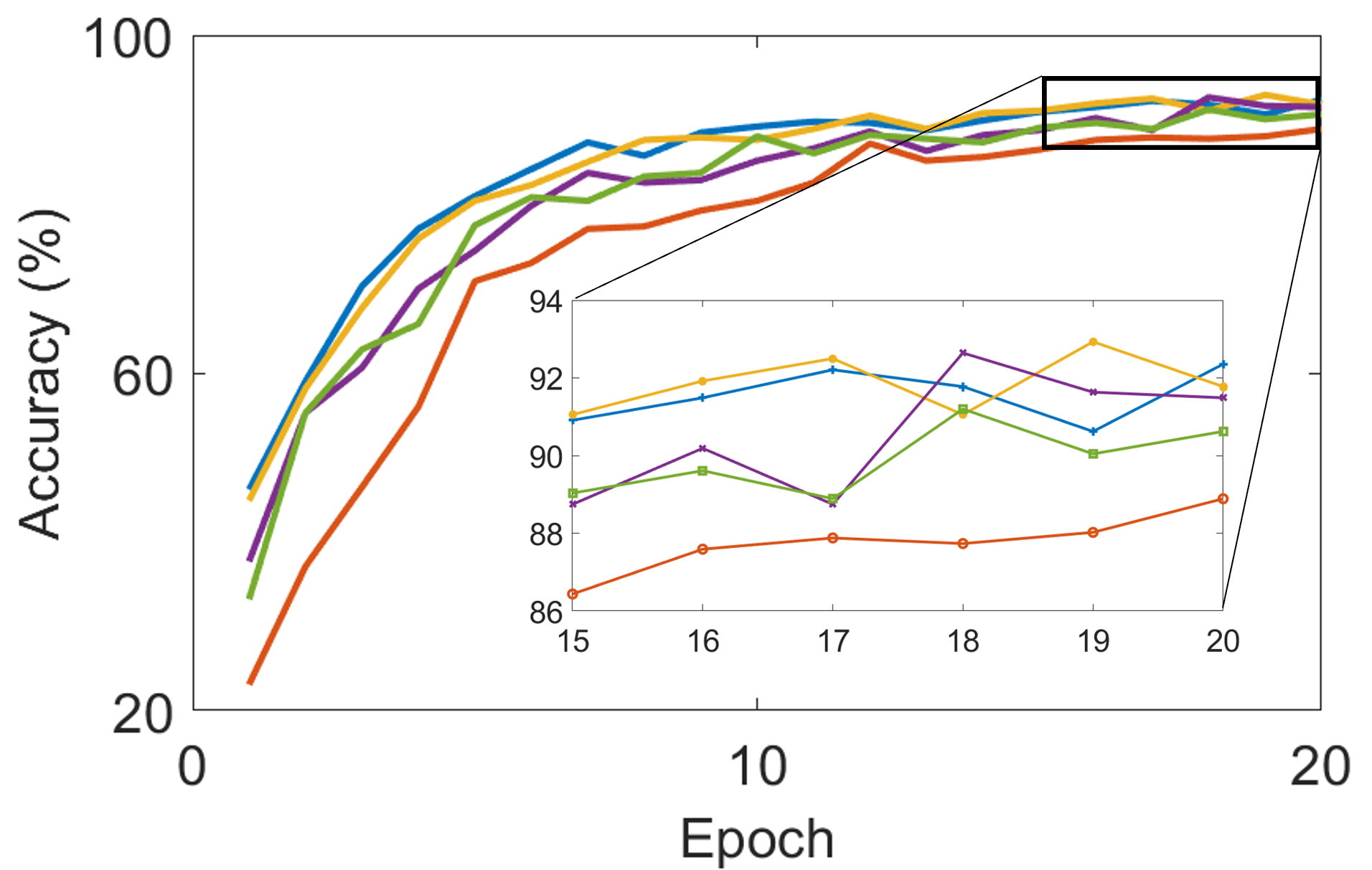}\captionsetup{justification=centering}
     \subcaption{Comp 2}
     \includegraphics[width=0.8\linewidth]{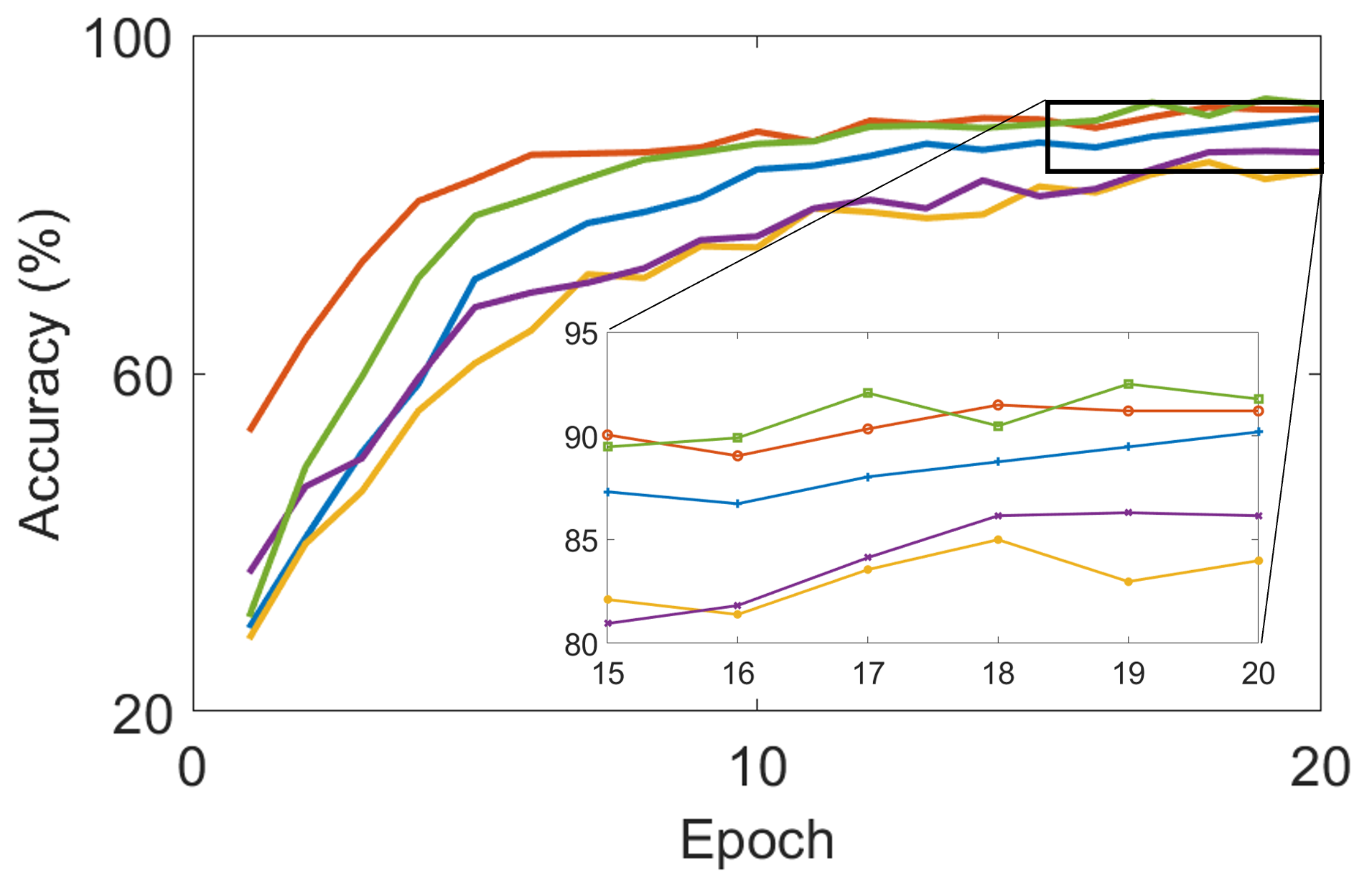}\captionsetup{justification=centering}
     \subcaption{Comp 3}
    \end{multicols}
\caption{Training results of each scheduling algorithm in VGG16.}
\label{fig:Training Result in VGG16}
\end{figure*}

\begin{figure*}[t]\centering
    \includegraphics[width=0.5\linewidth]{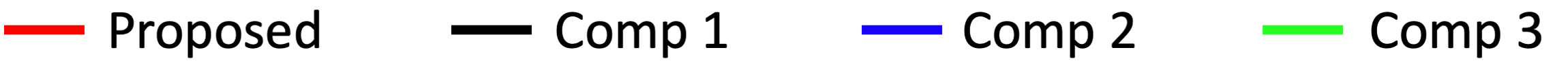}\captionsetup{justification=centering}\\
    \begin{multicols}{2}
     \includegraphics[width=0.8\linewidth]{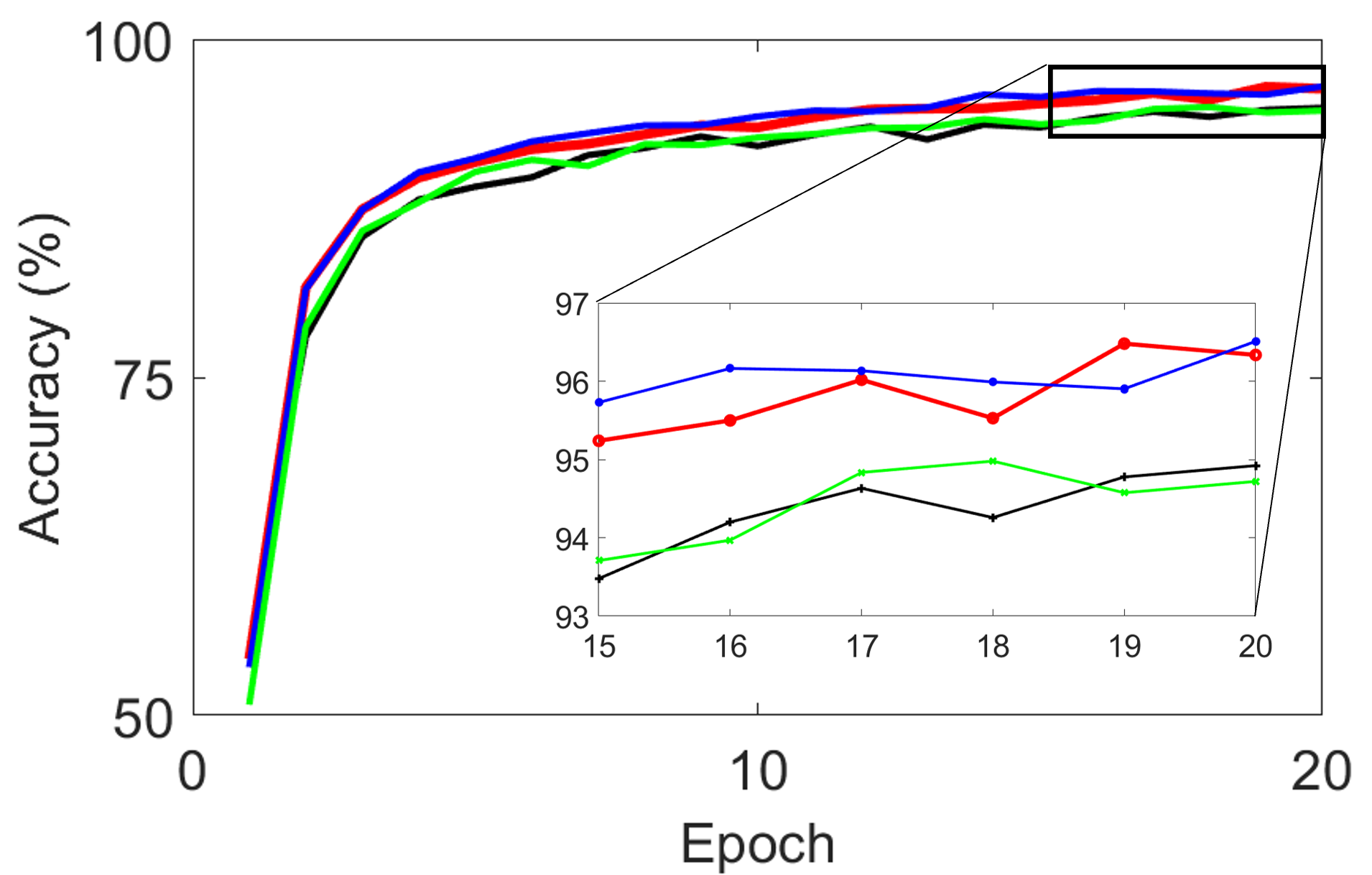}\captionsetup{justification=centering}
     \subcaption{Avg. accuracy}
     \includegraphics[width=0.8\linewidth]{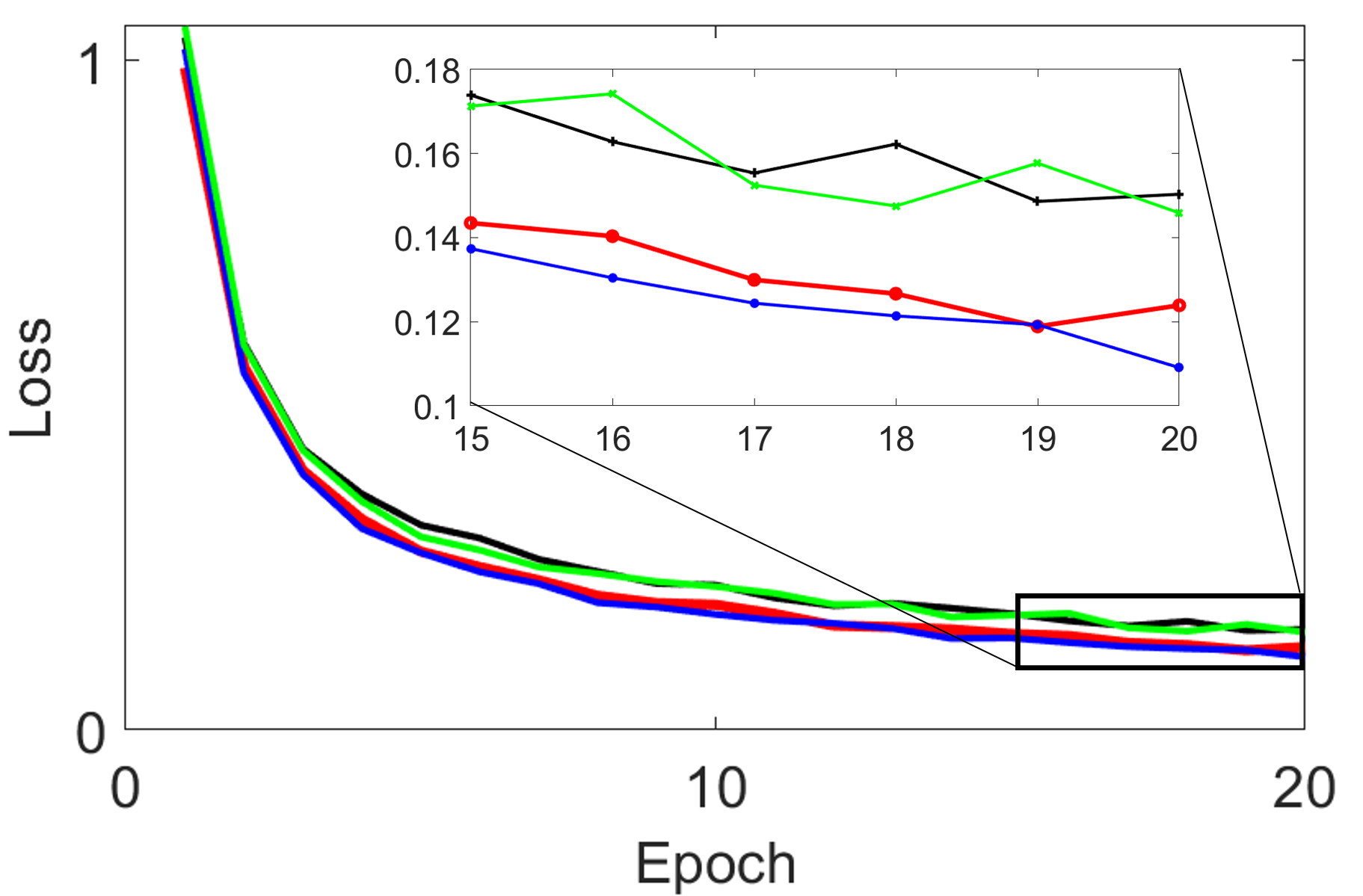}\captionsetup{justification=centering}
     \subcaption{Avg. loss}
     \end{multicols}
     
     \begin{multicols}{2}
     \includegraphics[width=0.8\linewidth]{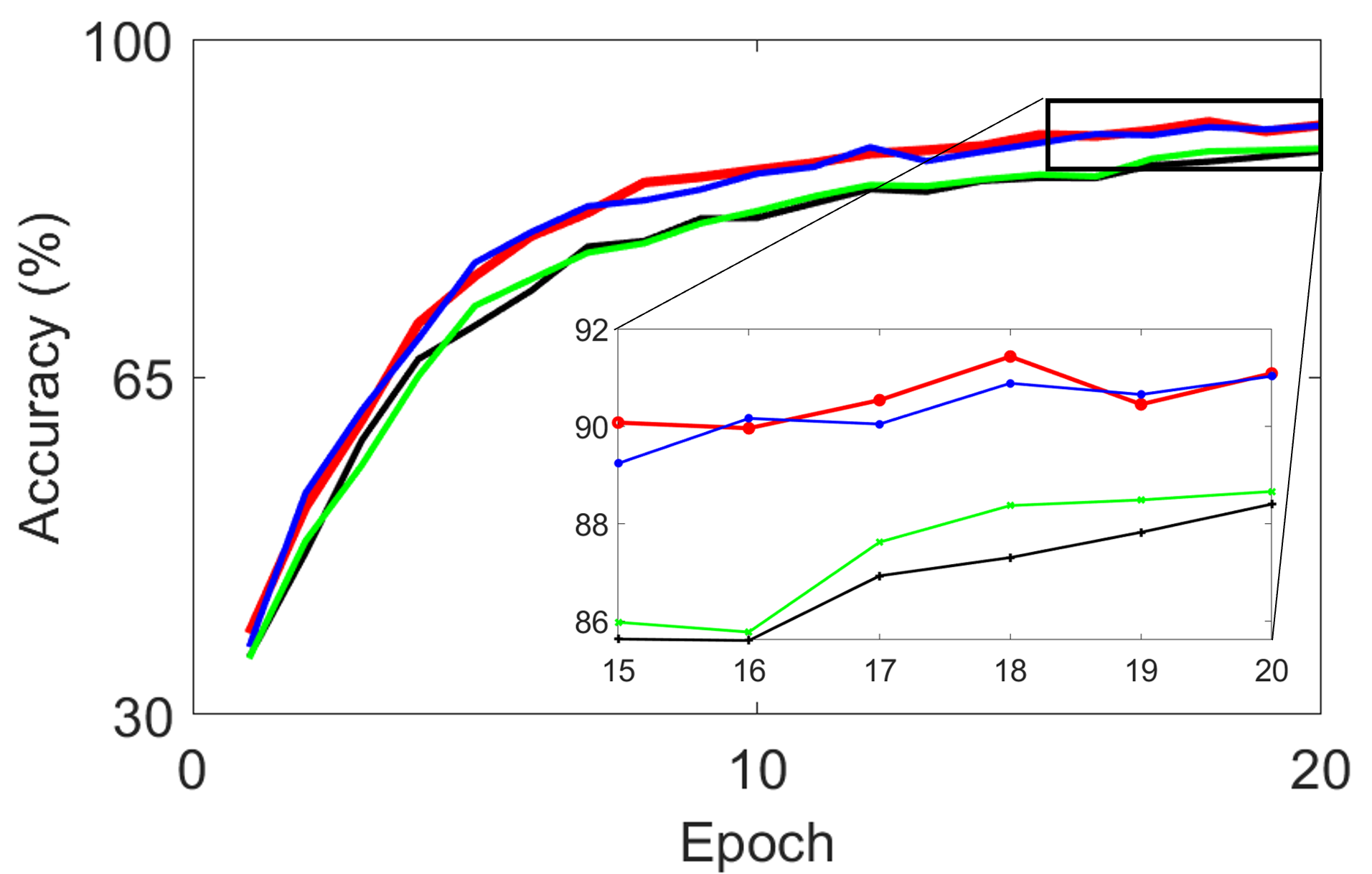}\captionsetup{justification=centering}
     \subcaption{Avg. accuracy}
     \includegraphics[width=0.8\linewidth]{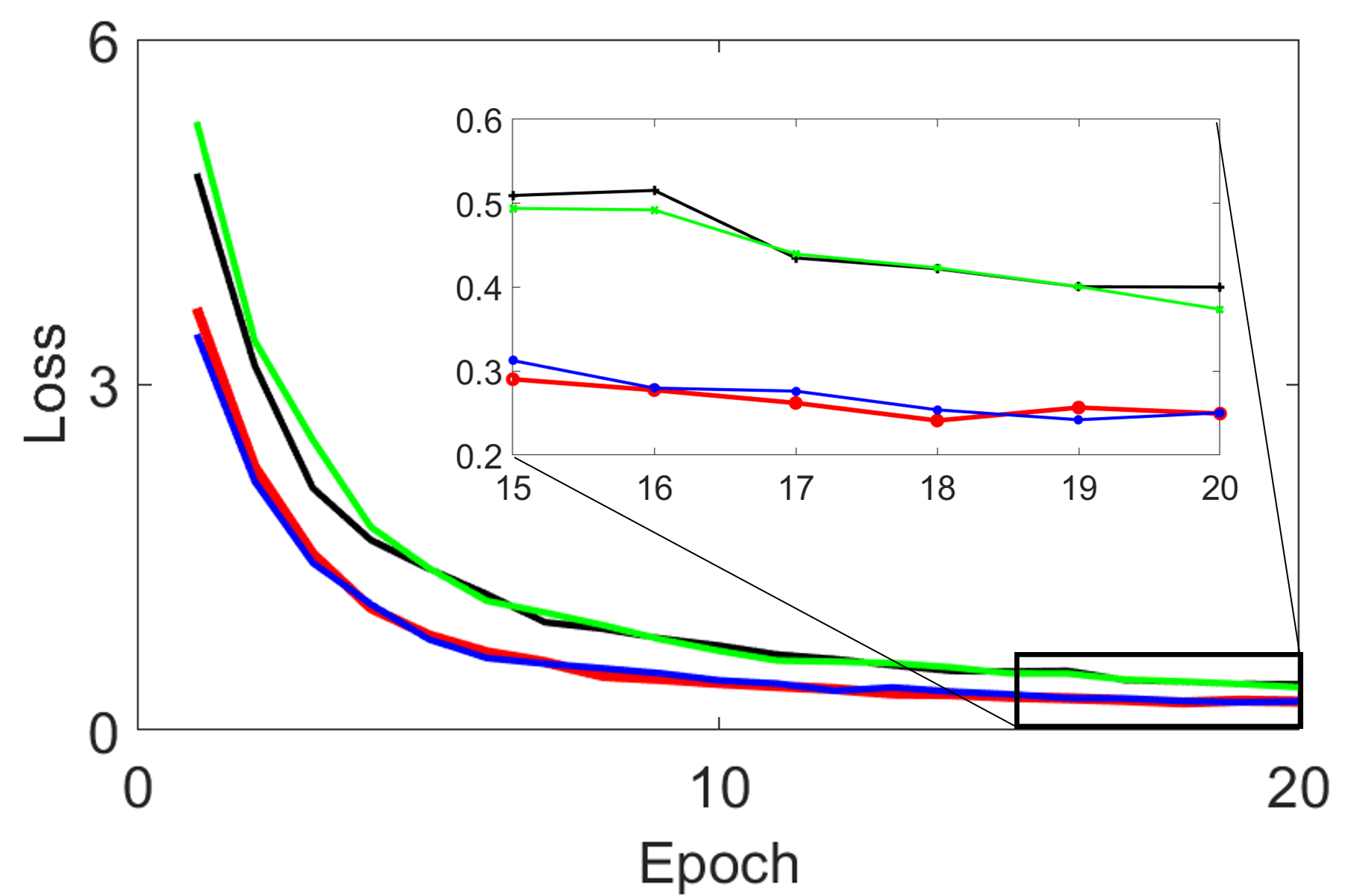}\captionsetup{justification=centering}
     \subcaption{Avg. loss}
    \end{multicols}
\caption{Total average results of each scheduling algorithm in both models.}
\label{fig:Total_Results}
\end{figure*}

\begin{table}[t]
\small
\caption{Training results of ResNet50 and VGG16 using the data collected by multi-UAVs.}
    \centering
        \begin{tabular}{c|rrrr}
        \toprule[1pt]
        Method & \textbf{Proposed} & Comp 1 & Comp 2 & Comp 3 \\ \midrule
        \multicolumn{5}{c}{\textbf{ResNet50}} \\ \midrule
        Tower \#1 & \textbf{96.4\,\%} & 95.4\,\% & 95.7\,\% & 95.1\,\% \\
        Tower \#2 & \textbf{96.3\,\%} & 95.2\,\% & 96.8\,\% & 95.5\,\% \\
        Tower \#3 & \textbf{96.0\,\%} & 94.5\,\% & 97.0\,\% & 93.2\,\% \\
        Tower \#4 & \textbf{96.3\,\%} & 94.6\,\% & 95.7\,\% & 93.5\,\% \\
        Tower \#5 & \textbf{96.8\,\%} & 94.8\,\% & 97.4\,\% & 96.2\,\% \\ \midrule
        Total Accuracy & \textbf{96.3\,\%} & 94.9\,\% & 96.5\,\% & 94.7\,\% \\ 
        Standard Deviation & \textbf{0.258} & 0.346 & 0.697 & 1.161 \\ 
        Total Loss & \textbf{0.124} & 0.150 & 0.109 & 0.146 \\ \midrule
        \multicolumn{5}{c}{\textbf{VGG16}} \\ \midrule
        Tower \#1 & \textbf{88.5\,\%} & 89.9\,\% & 92.4\,\% & 90.2\,\% \\
        Tower \#2 & \textbf{90.9\,\%} & 88.7\,\% & 88.9\,\% & 91.2\,\% \\
        Tower \#3 & \textbf{92.7\,\%} & 90.8\,\% & 91.8\,\% & 84.0\,\% \\
        Tower \#4 & \textbf{91.4\,\%} & 86.3\,\% & 91.5\,\% & 86.1\,\% \\
        Tower \#5 & \textbf{92.1\,\%} & 86.3\,\% & 90.6\,\% & 91.8\,\% \\ \midrule
        Total Accuracy & \textbf{91.1\,\%} & 88.4\,\% & 91.0\,\% & 88.6\,\% \\
        Standard Deviation & \textbf{1.446} & 1.84 & 1.218 & 3.064 \\
        Total Loss & \textbf{0.249} & 0.401 & 0.251 & 0.374 \\
        \bottomrule[1pt]
        \end{tabular}
        \label{tab:training_result_in_each_model}
\end{table}

\subsubsection{Learning Accuracy at Towers}\label{sec:4-2-3}
% Intro
Figs.~\ref{fig:training_result_in_ResNet50}--\ref{fig:Total_Results} and Table~\ref{tab:training_result_in_each_model} show the various aspects of training performance of all towers with collected training data by multi-UAVs via all scheduling algorithms (i.e., Proposed, Comp1, Comp2, and Comp3).
Because towers perform classification (i.e., object recognition for abnormal behavior detection) between the flame and non-flame cases using ResNet50 and VGG16, we compared learning accuracy and training loss over all epochs in all scheduling algorithmss to evaluate the towers' training performances.

In Fig.~\ref{fig:training_result_in_ResNet50} and Fig.~\ref{fig:Training Result in VGG16}, we compare the proposed scheduling algorithm's performance using the training accuracy convergence status in ResNet50 and VGG16. The graph shows that the proposed and Comp2 algorithms have the highest accuracy (ResNet50: between $95$--$98$, VGG16: between $88$--$92$). Note that the learning accuracy of all $5$ towers in the proposed algorithm converges to the most similar levels in ResNet50 with the lowest standard deviation of accuracy among all scheduling algorithms, depending on the purpose of the fair data collection for towers. In addition, the proposed algorithm's standard deviation of learning accuracy in VGG16 is the second lowest following the Comp2 algorithm.
The Comp1 algorithm shows that the training accuracy has a lower performance level than the other algorithms.
As illustrated in Fig.~\ref{fig:Total Data Size}, model training results are affected by the size of data acquired in the towers according to the scheduling algorithm. That is, the larger and more evenly the data is collected, the more stable and a high level of convergence can be guaranteed and obtained.
The performance evaluation results using the Comp3 algorithm show that certain towers have very high performance, while some have the lowest convergence value (ResNet50: $93.2\%$, VGG16: $84.0\%$) in all experimental results using each model.
Because the data was randomly transferred without any criteria, the training accuracy converging for each tower has the most significant deviation from the others.
These results are more pronounced when comparing the average accuracy values for the five towers by the algorithm in Fig.~\ref{fig:Total_Results}(a)/(c). Similar to the average accuracy, when we compare the mean loss value over time, the proposed and Comp1 algorithms have the lowest loss convergence as the average accuracy is high. 
The proposed scheduling algorithm, which considers both the data variance and the power status of UAVs, shows effective learning results that do not differ significantly compared to the algorithm that considers only the data characteristics.

In a nutshell, it can be shown that all towers in our proposed and Comp2 algorithms have higher classification accuracy, faster increase rate, and less training loss than the other two algorithms in both ResNet50 and VGG16 models because the UAVs in the proposed and Comp2 algorithms deliver the larger and more fair number of training data than the other algorithms. Note that both the ResNet50 and VGG16 models of the proposed algorithm achieve similar training performances with the Comp2 algorithm, notwithstanding UAVs' fewer data collection and more efficient energy management in the proposed algorithm compared to the Comp2 algorithm, as shown in Table~\ref{tab:training_result_in_each_model}. In addition, the value of the total loss in the proposed algorithm is the second lowest in ResNet50 and the lowest in VGG16. As a result, we validate the performance of our proposed scheduling algorithm in terms of the training accuracy, standard deviation, and training loss of all towers by comparing it with the others. Lastly, we can also investigate the effect of the reward function by comparing Comp1 and Comp2 algorithms.

\section{Concluding Remarks}\label{sec:5}
This paper proposes a novel workload-aware Markov decision process (MDP)-based scheduling decision policies for fair contents distribution in energy-limited infrastructure/tower-assisted multi-UAV surveillance networks. 
In the proposed MDP-based scheduling algorithm between multi-towers and multi-UAVs, the optimization of data and power resource exchange decisions considering the workload-fairness is essential for high-performance surveillance object recognition learning model accuracy in towers in order to avoid too much or less training data in each tower (for avoiding computation burden and overfitting). Furthermore, it is also essential to consider power/energy-efficiency in UAVs.
Therefore, the corresponding novel optimization framework is designed, and in turn, the solution approach based on MDP-based reinforcement learning is also proposed for discrete-time sequential decision making in time-varying optimization computation and pseudo-polynomial optimality. 
Our data-intensive simulation-based performance evaluation results under the consideration of various UAV mobility/trajectory models show that our proposed MDP-based scheduling decision algorithm achieves desired performance improvements in terms of power-charging efficiency at UAVs, data distribution fairness at towers, and learning accuracy at towers. 

\bibliographystyle{IEEEtran}
\bibliography{ref_access,ref_aimlab}

\begin{IEEEbiography}[{\includegraphics[width=1in,height=1.25in,clip,keepaspectratio]{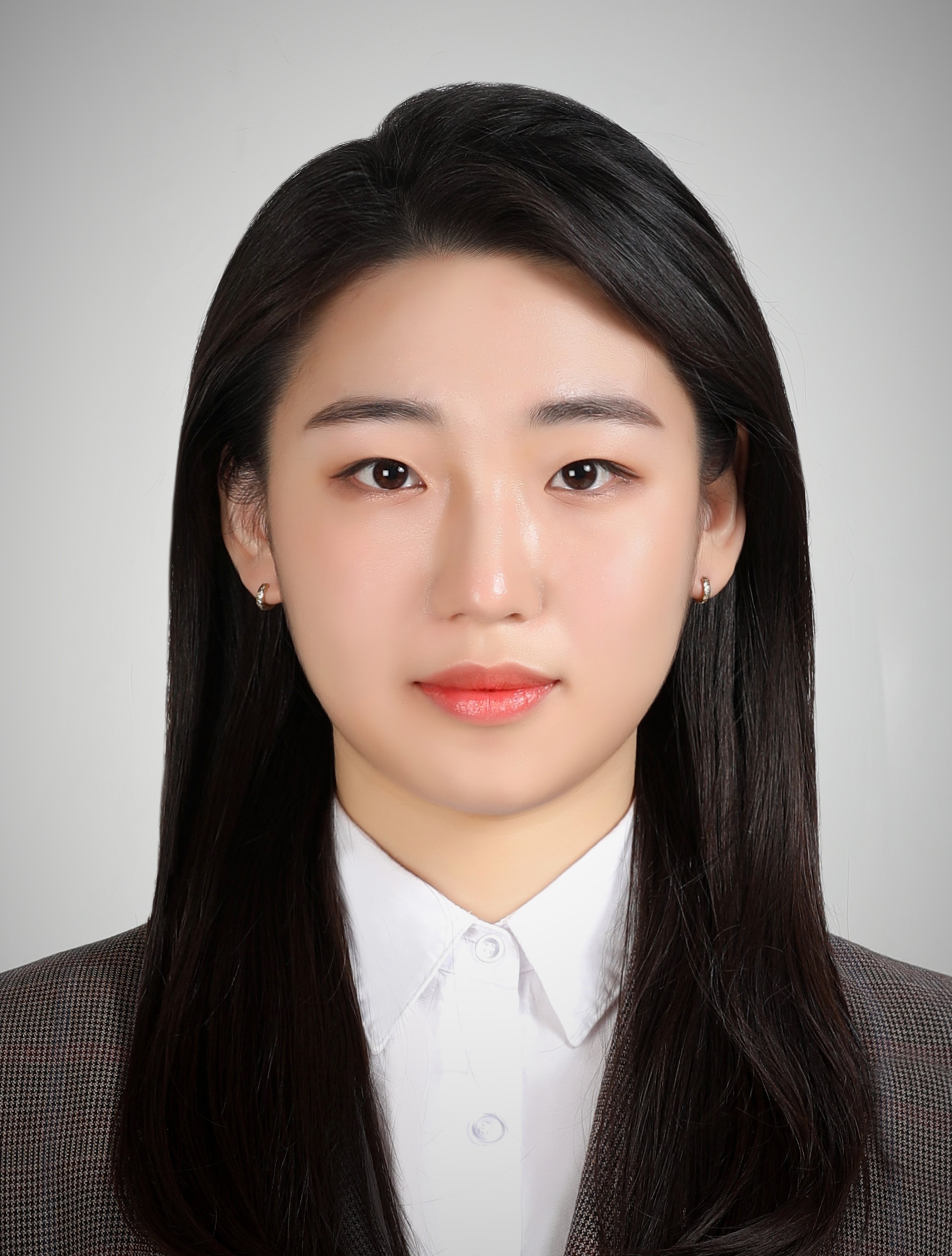}}]{Soohyun Park} is currently pursuing the Ph.D. degree in electrical and computer engineering at Korea University, Seoul, Republic of Korea. She received the B.S. degree in computer science and engineering from Chung-Ang University, Seoul, Republic of Korea, in 2019. %Her research focuses include deep learning algorithms and their applications. 

She was a recipient of the IEEE Vehicular Technology Society (VTS) Seoul Chapter Award in 2019.\end{IEEEbiography}

\begin{IEEEbiography}[{\includegraphics[width=1in,height=1.25in,clip,keepaspectratio]{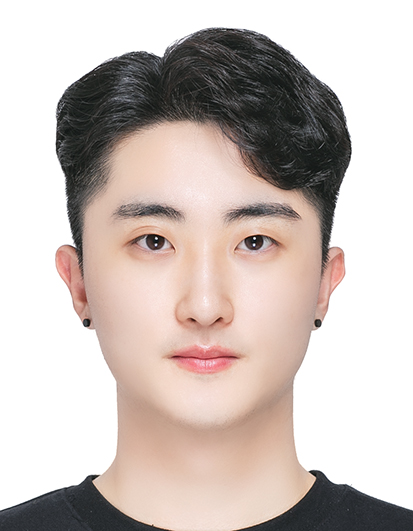}}]{Chanyoung Park} is currently a Ph.D. student in Electrical and Computer Engineering at Korea University, Seoul, Republic of Korea, since September 2022. He received the B.S. degree in electrical and computer engineering from Ajou University, Suwon, Republic of Korea, in 2022, with an honor (early graduation). His research focuses include deep learning algorithms and their applications to networks. 
\end{IEEEbiography}

\begin{IEEEbiography}[{\includegraphics[width=1in,height=1.25in,clip,keepaspectratio]{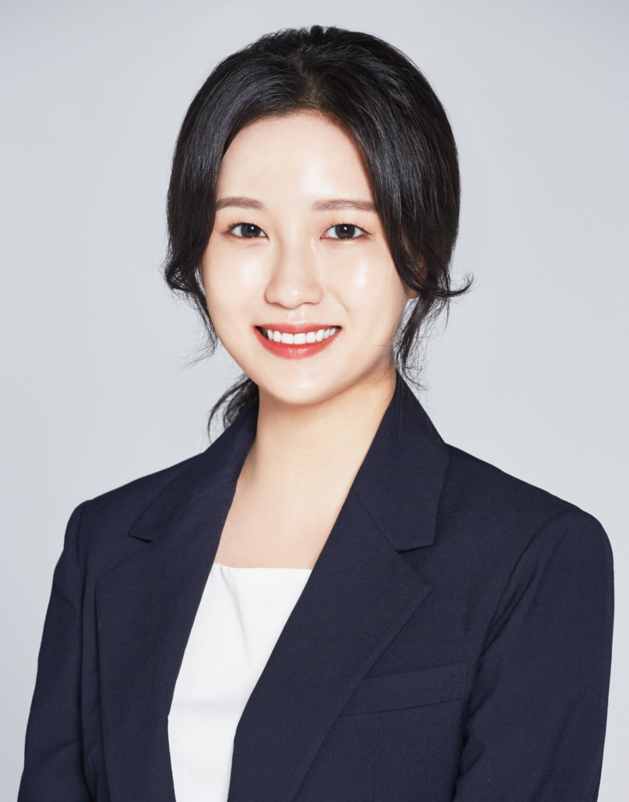}}]{Soyi Jung} has been an assistant professor at the Department of Electrical of Computer Engineering, Ajou University, Suwon, Republic of Korea, since September 2022. Before joining Ajou University, she was an assistant professor at Hallym University, Chuncheon, Republic of  Korea, from 2021 to 2022; a visiting scholar at Donald Bren School of Information and Computer Sciences, University of California, Irvine, CA, USA, from 2021 to 2022; a research professor at Korea University, Seoul, Republic of Korea, in 2021; and a researcher at Korea Testing and Research (KTR) Institute, Gwacheon, Republic of Korea, from 2015 to 2016. She received her B.S., M.S., and Ph.D. degrees in electrical and computer engineering from Ajou University, Suwon, Republic of Korea, in 2013, 2015, and 2021, respectively. Her current research interests include network optimization for autonomous vehicles communications, distributed system analysis, big-data processing platforms, and probabilistic access analysis. She was a recipient of Best Paper Award by KICS (2015), Young Women Researcher Award by WISET and KICS (2015), Bronze Paper Award from IEEE Seoul Section Student Paper Contest (2018), ICT Paper Contest Award by Electronic Times (2019), and IEEE ICOIN Best Paper Award (2021).
\end{IEEEbiography}

\begin{IEEEbiography}[{\includegraphics[width=1in,height=1.25in,clip,keepaspectratio]{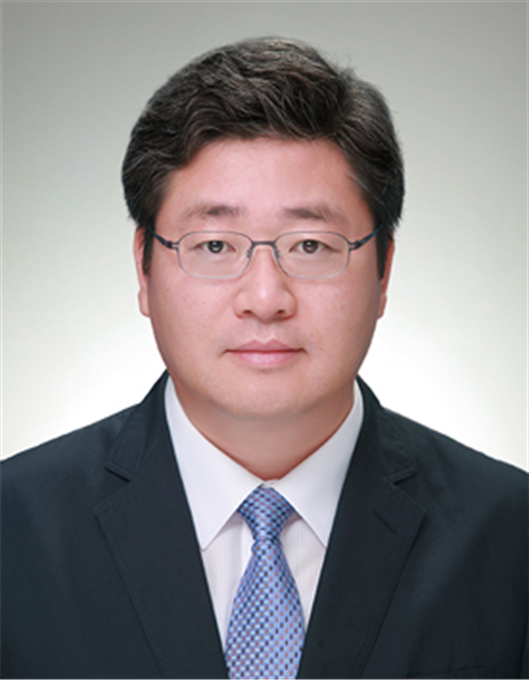}}]{Jae-Hyun Kim} received the B.S., M.S., and Ph.D. degrees, all in computer science and engineering, from Hanyang University, Ansan, Korea, in 1991, 1993, and 1996 respectively. In 1996, he was with the Communication Research Laboratory, Tokyo, Japan, as a Visiting Scholar. From April 1997 to October 1998, he was a postdoctoral fellow at the department of electrical engineering, University of California, Los Angeles. From November 1998 to February 2003, he worked as a member of technical staff in Performance Modeling and QoS management department, Bell laboratories, Lucent Technologies, Holmdel, NJ. He has been with the department of electrical and computer engineering, Ajou University, Suwon, Korea, as a professor since 2003. 

His research interests include medium access control protocols, QoS issues, cross layer optimization for wireless communication, and satellite communication. He is the Center Chief of Satellite Information Convergence Application Services Research Center (SICAS) sponsored by Institute for Information $\&$ Communications Technology Promotion in Korea. He is Chairman of the Smart City Committee of 5G Forum in Korea since 2018. He is Executive Director of the Korea Institute of Communication and Information Sciences (KICS). He is a member of the IEEE, KICS, the Institute of Electronics and Information Engineers (IEIE), and the Korean Institute of Information Scientists and Engineers (KIISE). 
\end{IEEEbiography}

\begin{IEEEbiography}[{\includegraphics[width=1in,height=1.25in,clip,keepaspectratio]{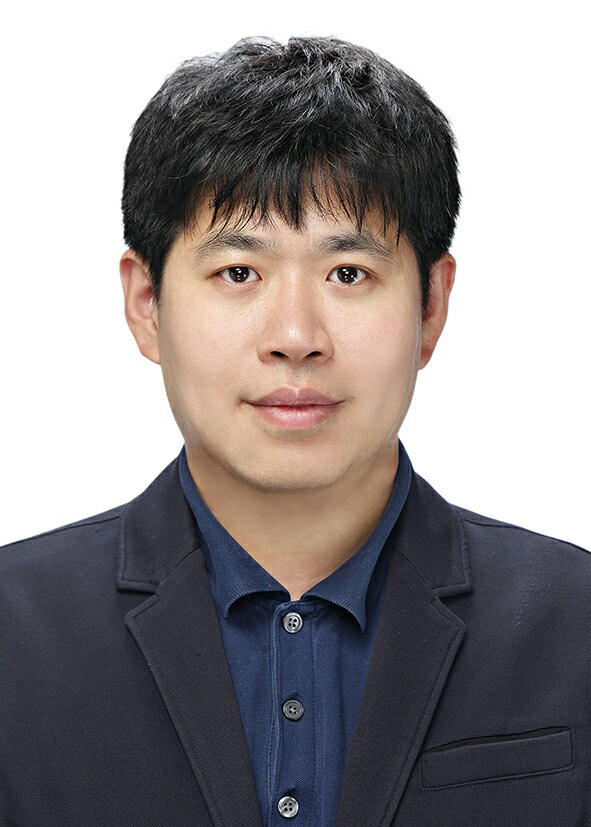}}]{Joongheon Kim} (M'06--SM'18) has been with Korea University, Seoul, Korea, since 2019, where he is currently an associate professor at the School of Electrical Engineering and also an adjunct professor at the Department of Communications Engineering (established/sponsored by Samsung Electronics) and the Department of Semiconductor Engineering (established/sponsored by SK Hynix). He received the B.S. and M.S. degrees in computer science and engineering from Korea University, Seoul, Korea, in 2004 and 2006; and the Ph.D. degree in computer science from the University of Southern California (USC), Los Angeles, CA, USA, in 2014. Before joining Korea University, he was a research engineer with LG Electronics (Seoul, Korea, 2006--2009), a systems engineer with Intel Corporation (Santa Clara, CA, USA, 2013--2016), and an assistant professor of computer science and engineering with Chung-Ang University (Seoul, Korea, 2016--2019). 

He serves as an editor for \textsc{IEEE Transactions on Vehicular Technology}, \textsc{IEEE Transactions on Machine Learning in Communications and Networking}, and \textsc{IEEE Communications Standards Magazine}. He is also a distinguished lecturer for \textit{IEEE Communications Society (ComSoc)} and \textit{IEEE Systems Council}.

He was a recipient of Annenberg Graduate Fellowship with his Ph.D. admission from USC (2009), Intel Corporation Next Generation and Standards (NGS) Division Recognition Award (2015), \textsc{IEEE Systems Journal} Best Paper Award (2020), IEEE ComSoc Multimedia Communications Technical Committee (MMTC) Outstanding Young Researcher Award (2020), IEEE ComSoc MMTC Best Journal Paper Award (2021), and Best Special Issue Guest Editor Award by \textit{ICT Express (Elsevier)} (2022). He also received several awards from IEEE conferences including IEEE ICOIN Best Paper Award (2021), IEEE Vehicular Technology Society (VTS) Seoul Chapter Awards (2019, 2021, and 2022), and IEEE ICTC Best Paper Award (2022). 
\end{IEEEbiography}

\EOD

\end{document}